\documentclass[fleqn,usenatbib]{mnras}

\usepackage{newtxtext,newtxmath}
\usepackage[T1]{fontenc}

\usepackage{graphicx}	% Including figure files
\usepackage{amsmath}	% Advanced maths commands
\usepackage{xcolor}
\usepackage[flushleft]{threeparttable}
\usepackage{verbatim}
\usepackage{array}
\usepackage{tabularx}
\usepackage{colortbl}
\usepackage{hyperref}
\usepackage{float}
\usepackage{subcaption}
\usepackage{booktabs}
\usepackage{adjustbox}
\usepackage{pdflscape}

\DeclareRobustCommand{\VAN}[3]{#2}
\let\VANthebibliography\thebibliography
\def\thebibliography{\DeclareRobustCommand{\VAN}[3]{##3}\VANthebibliography}

\newcommand{\kepler}{{\it Kepler}}

\newcommand{\NGTS}{NGTS}
\newcommand{\TESS}{{\it TESS}}

\newcommand{\Spitzer}{\textit{Spitzer}}
\newcommand{\HARPS}{HARPS}

\newcommand{\kms}{km\,s$^{-1}$}
\newcommand{\ms}{m\,s$^{-1}$}

\newcommand{\mstar}{\mbox{M$_{*}$}}
\newcommand{\rstar}{\mbox{R$_{*}$}}

\newcommand{\msun}{\mbox{M$_{\odot}$}}
\newcommand{\rsun}{\mbox{R$_{\odot}$}}
\newcommand{\rearth}{R$_{\oplus}$}
\newcommand{\mearth}{M$_{\oplus}$}
\newcommand{\gccc}{g\,cm$^{-3}$}

\newcommand{\teff}{$T_{\rm eff}$}

\newcommand{\logg}{$\log g$}
\newcommand{\feh}{[Fe/H]}
\newcommand{\vmac}{$v_{\rm mac}$}

\newcommand{\vsini}{$v \sin i_\star$}   
 
\newcommand{\vmic}{$V_{\rm mic}$}

\newcommand{\halpha}{H$\alpha$}                   %H I recombination lines 

\newcommand{\cai}{Ca\,{\sc I} }

\newcommand{\fei}{[Fe\,{\sc I}] }

\newcommand{\mgi}{Mg\,{\sc I} }

\newcommand{\gc}{g~cm$^{-3}$}
\title[TOI-431]{TOI-431/HIP 26013: a super-Earth and a sub-Neptune transiting a bright, early K dwarf, with a third RV planet}
\author[A. Osborn et al.]{\parbox{\textwidth}{\Large
Ares~Osborn$^{1,2}$,
David~J.~Armstrong$^{1,2}$, 
Bryson~Cale$^{3}$,                  %primary iSHELL contributor
Rafael~Brahm$^{4,5}$,               %primary FEROS contributor
Robert~A.~Wittenmyer$^{6}$,         %primary MINERVA contributor
Fei~Dai$^{7}$,
Ian~J.~M.~Crossfield$^{8}$,          %primary Spitzer contributor -- ianc@ku.edu
Edward~M.~Bryant$^{1,2}$,              %primary NGTS contributor
Vardan~Adibekyan$^{9,10}$, 
Ryan~Cloutier$^{11,12}$
% ...
}
\vspace{0.2cm}
\\
\parbox{\textwidth}{
The authors' affiliations are shown in Appendix \ref{sec:affiliations}.\\
*E-mail: e.osborn@warwick.ac.uk}\vspace{-0.3cm}}
\date{\vspace{-0.5cm}Accepted XXX. Received YYY; in original form ZZZ}

\pubyear{2020}

\begin{document}
\label{firstpage}
\pagerange{\pageref{firstpage}--\pageref{lastpage}}
\maketitle

%%%%%%%%%%%%%%%%%%% ABSTRACT %%%%%%%%%%%%%%%%%%%%

\begin{abstract}
We present the bright (V$_{mag} = 9.12$), multi-planet system TOI-431, characterised with photometry and radial velocities. We estimate the stellar rotation period to be $30.5 \pm 0.7$ days using archival photometry and radial velocities.
TOI-431\,b is a super-Earth with a period of 0.49\,days, a radius of 1.28 $\pm$ 0.04\,\rearth, a mass of $3.07 \pm 0.35$\,\mearth, and a density of $8.0 \pm 1.0$\,\gc; TOI-431\,d is a sub-Neptune with a period of 12.46\,days, a radius of $3.29 \pm 0.09$\,\rearth, a mass of $9.90^{+1.53}_{-1.49}$\,\mearth, and a density of $1.36 \pm 0.25$\,\gc. We find a third planet, TOI-431\,c, in the \HARPS\ radial velocity data, but it is not seen to transit in the \TESS\ light curves. It has an $M \sin i$ of $2.83^{+0.41}_{-0.34}$\,\mearth, and a period of 4.85\,days.
TOI-431\,d likely has an extended atmosphere and is one of the most well-suited \TESS\ discoveries for atmospheric characterisation, while the super-Earth TOI-431\,b may be a stripped core. These planets straddle the radius gap, presenting an interesting case-study for atmospheric evolution, and TOI-431\,b is a prime \TESS\ discovery for the study of rocky planet phase curves.
\end{abstract}


%%%%%%%%%%%%%%%%%%%%%%%%%%%%%%%%%%%%%%%%%%%%%%%%%%

%%%%%%%%%%%%%%%%% BODY OF PAPER %%%%%%%%%%%%%%%%%%

\section{Introduction} \label{sec:intro}
The discoveries of the \kepler\ Space Telescope \citep{Borucki2010} provided us with the means to make statistical studies on the exoplanet population for the first time: \kepler\ has shown us that Neptune-sized planets are more common than large gas giants \citep{Fressin2013}, and that super-Earths are the most abundant planet type \citep{Petigura2013}. It became possible to look for trends that might elucidate planetary formation mechanisms; one such trend discovered is a bi-modality in the radius distribution of small planets. Often dubbed the ``photoevaporation valley,'' the commonly posited explanation for its existence is photoevaporation of close-in planetary atmospheres \citep{Owen2017,Fulton2017,Fulton2018,VanEylen2018,Cloutier2020a}. Planets above the radius gap have retained gaseous envelopes, while planets below are theorised to have been stripped of any gas to become naked cores. Multi-planet systems have been discovered containing planets that lie both below and above the radius gap \citep[e.g.][]{Gunther2019,Cloutier2020-LTT3780}, and such systems are important when considering how evolution mechanisms may sculpt the radius gap as they allow testing of atmospheric evaporation and bulk composition models. 

Further to the discovery of the radius gap, a paucity of intermediate-sized planets at short periods ($\leq$ 3\,days) dubbed the ``Neptune/sub-Jovian Desert'' \citep{Szabo2011,Beauge2013,Helled2016,Lundkvist2016,Mazeh2016,Owen2018}, can be seen in both the mass-period and radius-period distribution of exoplanets, and \citet{Mazeh2016} and \citet{Owen2018} derived boundaries for this triangular-shaped region, and the potential mechanisms behind their existence. 

The Transiting Exoplanet Survey Satellite \citep[\textit{TESS},][]{Ricker2015} is now building upon the legacy of \kepler. Unlike \kepler, \TESS\ has been optimised to look at bright stars, enabling high precision radial velocity follow up of planetary candidates to determine their masses, and additional follow-up (with the James Webb Space Telescope, JWST, for example) will allow us to study their atmospheres. Over the course of its two year primary mission, which came to an end in July 2020, over 2000 \TESS\ Objects of Interest (TOIs) were released, and there have been many discoveries that contribute to fulfilling its Level-1 mission goal to measure the masses and radii of at least 50 planets with radii smaller than 4 \rearth\ \citep[e.g.][]{Huang2018, Gandolfi2018, Cloutier2019, Dragomir2019, Dumusque2019, Luque2019, Diaz2020, Astudillo2020, Cloutier2020-TOI1235, Cloutier2020-LTT3780, Nielsen2020, Armstrong2020}. 

We present here the discovery of TOI-431\,b, c, and d. TOI-431\,b and d are a super-Earth and sub-Neptune respectively, discovered first by \TESS\ and confirmed via extensive follow up: high-precision Doppler spectroscopy from the High Accuracy Radial velocity Planet Searcher \citep[HARPS,][]{Pepe2002} and the HIgh REsolution Spectrograph \citep[HIRES,][]{Vogt1994} which allows us to determine their masses; additional Doppler spectroscopy from iSHELL \citep{Rayner2016}, FEROS \citep{Kaufer1998}, and M\textsc{inerva}-Australis \citep{Addison2019}; ground-based transit detections from NGTS \citep{wheatley18ngts} and the LCOGT\,1m network \citep{Brown2013}; and a double-transit from the Spitzer space telescope. Both TOI-431\,b and d contribute to the \TESS\ Level-1 mission goal. TOI-431\,c is an additional planet that we have found in the \HARPS\ radial velocity data, and it is not seen to transit.
We describe the observations made and the stellar analysis of the TOI-431 system in Section \ref{sec:obs}; our joint-fit model of the system in Section \ref{sec:jointfit}; and put this system into context in Section \ref{sec:disc}.

\section{Observations} \label{sec:obs}
\begin{center} % TO BE COMPLETED
\begin{table}
    \caption{Details of the TOI-431 system.}
    \label{tab:toi-431}
    \begin{threeparttable}
    \begin{tabularx}{\columnwidth}{ l l X }
    \hline
    \textbf{Property} & \textbf{Value} & \textbf{Source} \\
    \hline
    \textbf{Identifiers}                & & \\
    TIC ID      & 31374837              & TICv8 \\
    HIP ID      & 26013                 &   \\
    2MASS ID    & 05330459-2643286      & 2MASS \\
    Gaia ID     & 2908664557091200768   & GAIA EDR3 \\
    & & \\
    \textbf{Astrometric properties}     & & \\
    R.A. (J2016.0)  & 05:33:04.62           & GAIA EDR3 \\
    Dec (J2016.0)   & -26:43:25.86          & GAIA EDR3 \\
    Parallax (mas)  & 30.65 $\pm$ 0.01      & GAIA EDR3 \\
    Distance (pc)   & 32.61 $\pm$ 0.01      & \citet{Bailer-Jones2021} \\
    $\mu_{\rm{R.A.}}$ (mas yr$^{-1}$)       & 16.89 $\pm$ 0.01   & GAIA EDR3 \\
    $\mu_{\rm{Dec}}$ (mas yr$^{-1}$)        & 150.78 $\pm$ 0.01  & GAIA EDR3 \\
    & & \\
    \textbf{Photometric properties} & & \\
    TESS (mag)  & 8.171 $\pm$ 0.006    & TICv8 \\
    B (mag)     & 10.10 $\pm$ 0.03     & TICv8 \\
    V (mag)     & 9.12 $\pm$ 0.03      & TICv8 \\
    G (mag)     & 8.7987 $\pm$ 0.0003  & GAIA EDR3 \\
    J (mag)     & 7.31 $\pm$ 0.03      & 2MASS \\
    H (mag)     & 6.85 $\pm$ 0.03      & 2MASS \\
    K (mag)     & 6.72 $\pm$ 0.02      & 2MASS \\
    \hline
    \end{tabularx}
    \begin{tablenotes}
    \item Sources: TICv8 \citep{Stassun2019}, 2MASS \citep{Skrutskie2006}, Gaia Early Data Release 3 \citep{GAIA_DR3}
    \end{tablenotes}
    \end{threeparttable}
\end{table}
\end{center}

\subsection{Photometry}

\subsubsection{TESS photometry} % TO BE FILLED
\label{TESSphotom}

The TOI-431 system (TIC\,31374837, HIP\,26013) was observed in \TESS\ Sectors 5 (Nov 15 to Dec 11 2018) and 6 (Dec 15 2018 to Jan 6 2019) on Camera 2 in the 2-minute cadence mode ($t_{\rm exp} = 2$\,min).
TOI-431.01 (now TOI-431\,d) was flagged on Feb 8 2019 by the MIT Quick-Look Pipeline \citep[QLP,][]{Huang2019} with a signal-to-noise ratio (SNR) of 58; the Sector 5 light curve reveals 2 deep transits of TOI-431\,d, but further transits of this planet fell in the data gaps in S6. TOI-431\,d passed all Data Validation tests \citep[see][]{Twicken:DVdiagnostics2018} and model fitting \citep[see][]{Li:DVmodelFit2019}; additionally, the difference image centroiding results place the transit signature source within $\sim 3$ arcsec of the target star. TOI-431.02 (now TOI-431\,b) was flagged later, on June 6,  after identification by the TESS Science Processing Operations Center (SPOC) pipeline \citep{Jenkins2016} with an SNR of 24 in a combined transit search of Sectors 5-6. 

We used the publicly available photometry provided by the SPOC pipeline, and used the Presearch Data Conditioning Simple Aperture Photometry (\textsc{PDCSAP\textunderscore FLUX}), which has common trends and artefacts removed by the SPOC Presearch Data Conditioning (PDC) algorithm \citep{Twicken2010,Smith2012,Stumpe2012,Stumpe2014}. The median-normalised PDCSAP flux, without any further detrending, is shown in the top panel of Fig. \ref{fig:TESS}. 

\subsubsection{LCOGT photometry} % Ryan Cloutier
\label{LCOphotom}
To confirm the transit timing and depth, and to rule out a nearby eclipsing binary (NEB) as the source of the \TESS\ transit events, we obtained three seeing-limited transit observations of TOI-431\,d in the $zs$-band. The light curves were obtained using the 1-m telescopes at the Cerro  Tololo Inter-American Observatory (CTIO) and the Siding Springs Observatory (SSO) as part of the Las Cumbres Observatory Global Telescope network \citep[LCOGT;][]{Brown2013}. Both telescopes are equipped with a $4096\times 4096$ Sinistro camera with a fine pixel scale of $0.39''$ pixel$^{-1}$.

We calibrated each sequence of images using the standard LCOGT \texttt{BANZAI} pipeline \citep{McCully2018}. The observations were scheduled using the \TESS\ Transit Finder, a customised version of the \texttt{Tapir} software package \citep{Jensen2013}. The differential light curves of TOI-431, and seven neighbouring sources within $2.5'$ based on the Gaia DR2 \citep{GAIA_DR2}, were derived from uncontaminated apertures using AstroImageJ \citep[AIJ;][]{Collins2017}. Two partial transits were obtained on UT December 9 2019 which covered the ingress and egress events from CTIO and SSO respectively (Figure~\ref{fig:LCO-NGTS}). We then obtained a second ingress observation on January 3 2020 from CTIO. Within each light curve, we detected the partial transit event on-target and cleared the field of NEBs down to $\Delta zs=6.88$\,mag.

\subsubsection{PEST photometry} % Karen Collins
\label{PESTphotom}
We also obtained a seeing-limited observation during the time of transit of TOI-431\,d on UT February 13 2020 using the Perth Exoplanet Survey Telescope (PEST) near Perth, Australia. The 0.3\,m telescope is equipped with a $1530\times1020$ SBIG ST-8XME camera with an image scale of 1$\farcs$2 pixel$^{-1}$, resulting in a $31\arcmin\times21\arcmin$ field of view. A custom pipeline based on {\tt C-Munipack}\footnote{http://c-munipack.sourceforge.net} was used to calibrate the images and extract the differential photometry, using an aperture with radius 6$\farcs$2. The images have typical stellar point spread functions (PSFs) with a FWHM of $\sim5\arcsec$.  Because the transit depth of TOI-431\,d is too shallow to detect from the ground with PEST, the target star was intentionally saturated to check the fainter nearby stars for possible NEBs that could be blended in the \TESS\ aperture. The data rule out NEBs in all 17 stars within $2\farcm5$ of the target star that are bright enough (\TESS\ magnitude < 17.4) to cause the \TESS\ detection of TOI-431\,d.

\subsubsection{Spitzer photometry} % Ian Crossfield
\label{Spitzerphotom}

Shortly after TOI-431 was identified and announced as a \TESS\ planet candidate, we identified TOI-431\,d as an especially interesting target for atmospheric characterization via transmission spectroscopy. We therefore scheduled one transit observation with the Spitzer Space Telescope to further refine the transit ephemeris and allow efficient scheduling of future planetary transits. We observed the system as part of Spitzer GO 14084 \citep{crossfield:2018spitzer} using the 4.5$\mu$m channel of the IRAC instrument \citep{fazio:2005}. We observed in subarray mode, which acquired 985 sets of 64 subarray frames, each with 0.4\,s integration time. These transit observations spanned UT times from May 23 2019 21:13 to May 24 2019 04:42, and were preceded and followed by shorter integrations observed off-target to check for bad or hot pixels. Our transit observations used Spitzer/IRAC in PCRS Peak-up mode to place the star as closely as possible to the well-characterized ``sweet spot'' on the IRAC2 detector.

\subsubsection{NGTS photometry} % Ed Bryant
\label{NGTSphotom}

The Next Generation Transit Survey \citep[NGTS;][]{wheatley18ngts} is an exoplanet hunting facility which consists of twelve 20\,cm diameter robotic telescopes and is situated at ESO's Paranal Observatory. Each NGTS telescope has a wide field-of-view of 8 square degrees and a plate scale of 5\,arcsec\, pixel$^{-1}$. NGTS observations are also afforded sub-pixel level guiding through the DONUTS auto-guiding algorithm \citep{mccormac13donuts}. A transit event of TOI-431\,d was observed using 5 NGTS telescopes on February 20 2020. On this night, a total of 5922 images were taken across the 5 telescopes, with each telescope observing with the custom NGTS filter and an exposure time of 10\,seconds. The dominant photometric noise sources in NGTS light curves of bright stars are Gaussian and uncorrelated between the individual telescope systems \citep{smith20multicam, bryant20multicam}. As such, we can use simultaneous observations with multiple NGTS telescopes to obtain high precision light curves.

All the NGTS data for TOI-431 were reduced using a custom aperture photometry pipeline which uses the SEP library for both source extraction and photometry \citep{bertin96sextractor, Barbary2016}. Bias, dark and flat field image corrections are found to not improve the photometric precision achieved, and so we do not apply these corrections during the image reduction. SEP and GAIA \citep{GAIA, GAIA_DR2} are both used to identify and rank comparison stars in terms of their brightness, colour, and CCD position relative to TOI-431 \citep[for more details on the photometry, see][]{bryant20multicam}.

%%%%%%%%%%%%%%%%%%%%%%%%%%%%%%%%%%%%%%%%%%%%%%%%%%%%%%%%%%%%%%%%%%%%%%%%

\subsection{Spectroscopy}

\subsubsection{HARPS high-resolution spectroscopy}
\label{HARPSspec}

TOI-431 was observed between February 2 and October 21 2019 with the High Accuracy Radial velocity Planet Searcher (\HARPS) spectrograph mounted on the ESO 3.6\,m telescope at the La Silla Observatory in Chile \citep{Pepe2002}. A total of 124 spectra were obtained under the NCORES large programme (ID 1102.C-0249, PI: Armstrong). The instrument (with resolving power $R = 115,000$) was used in high-accuracy mode (HAM), with an exposure time of 900\,s. Between 1 and 3 observations of the star were made per night. The standard offline HARPS data reduction pipeline was used to reduce the data, and a K5 template was used in a weighted cross-correlation function (CCF) to determine the radial velocities (RVs). Each epoch has further calculation of the bisector span (BIS), full-width at half-maximum (FWHM), and contrast of the CCF. This data is presented in Table \ref{tab:dataharps}.

In addition to this, there are 50 publicly available archival HARPS spectra dating from 2004 to 2015. 

\subsubsection{HIRES high-resolution spectroscopy}
\label{HIRESspec}

We obtained 28 high-resolution spectra of TOI-431 on the High Resolution Echelle Spectrometer of the 10m Keck I telescope \citep[Keck/HIRES,][]{Vogt1994}. The observation spans a temporal baseline from November 11 2019 to September 27 2020. We obtained an iodine-free spectrum on November 8 2019 as the template for radial velocity extraction. All other spectra were obtained with the iodine cell in the light path for wavelength calibration and line profile modeling. Each of these spectra were exposed for 4-8 min achieving a median SNR of 200 per reduced pixel near 5500\,Å. The spectra were analyzed with the forward-modelling Doppler pipeline described in \citet{Howard2010} for RV extraction. We analyzed the Ca II H \& K lines and extracted the $S_{\rm HK}$ using the method of \citet{Isaacson2010}. This data is presented in Table \ref{tab:datahires}.

\subsubsection{iSHELL spectroscopy} % Bryson Cale (sup. Peter Plavchan)
\label{iSHELLspec}

We obtained 108 spectra of TOI-431 during 11 nights with the iSHELL spectrometer on the {\em NASA Infrared Telescope Facility} \citep[IRTF,][]{2016SPIE.9908E..84R}, spanning 108 days from September-December 2019 . The exposure times were 5\,minutes, repeated 3-14 times within a night to reach a cumulative photon signal-to-noise ratio per spectral pixel varying from 131--334 at $\sim 2.4\,\mu$m (the approximate centre of the blaze for the middle order). This achieves a per-night RV precision of 3--8\,ms$^{-1}$ with a median of 5\,ms$^{-1}$. Spectra were reduced and RVs extracted using the methods outlined in \citet{caleetal2019}. 

\subsubsection{FEROS spectroscopy}
\label{FEROSspec}
TOI-431 was monitored with the Fiberfed Extended Range Optical Spectrograph \citep[FEROS,][]{kaufer:99}, installed on the MPG2.2 m telescope at La Silla Observatory, Chile. These observations were obtained in the context of the Warm gIaNts with tEss (WINE) collaboration, which focuses on the systematic characterization of \textit{TESS} transiting warm giant planets \citep[e.g.,][]{hd1397,jordan:2020}. FEROS has a spectral resolution of R $\approx48\,000$ and uses a comparison fibre that can be pointed to the background sky or illuminated by a Thorium-Argon lamp simultaneously with the execution of the science exposure. We obtained 10 spectra of TOI-431 between February 28 and March 12 2020. We used the simultaneous calibration technique to trace instrumental radial velocity variations, and adopted an exposure time of 300\,s, which translated in spectra with a typical signal-to-noise ratio per resolution element of 170. FEROS data was processed with the \texttt{ceres} pipeline \citep{brahm:2017:ceres}, which delivers precision radial velocity and line bisector span measurements through the cross-correlation technique. The cross-correlation was executed with a binary mask reassembling the properties of a G2-type dwarf star.

\subsubsection{M\textsc{inerva}-Australis spectroscopy}
\label{MAspec}

M\textsc{inerva}-Australis is an array of four PlaneWave CDK700 telescopes located in Queensland, Australia, fully dedicated to the precise radial-velocity follow-up of TESS candidates.  The four telescopes can be simultaneously fiber-fed to a single KiwiSpec R4-100 high-resolution (R=80,000) spectrograph \citep{barnes12,addison19,TOI257}.  TOI-431 was observed by M\textsc{inerva}-Australis in its early operations, with a single telescope, for 16 epochs between 2019 Feb 12 and 2019 April 17.  Each epoch consists of two 30-minute exposures, and the resulting radial velocities are binned to a single point. Radial velocities for the observations are derived for each telescope by cross-correlation, where the template being matched is the mean spectrum of each telescope.  The instrumental variations are corrected by using simultaneous Thorium-Argon arc lamp observations.

%%%%%%%%%%%%%%%%%%%%%%%%%%%%%%%%%%%%%%%%%%%%%%%%%%%%%%%%%%%%%%%%%%%%%%%%

\subsection{High resolution imaging} \label{sec:hri}

\begin{figure*} 
    \centering
    \begin{subfigure}{0.44\textwidth}
    \includegraphics[width=\columnwidth]{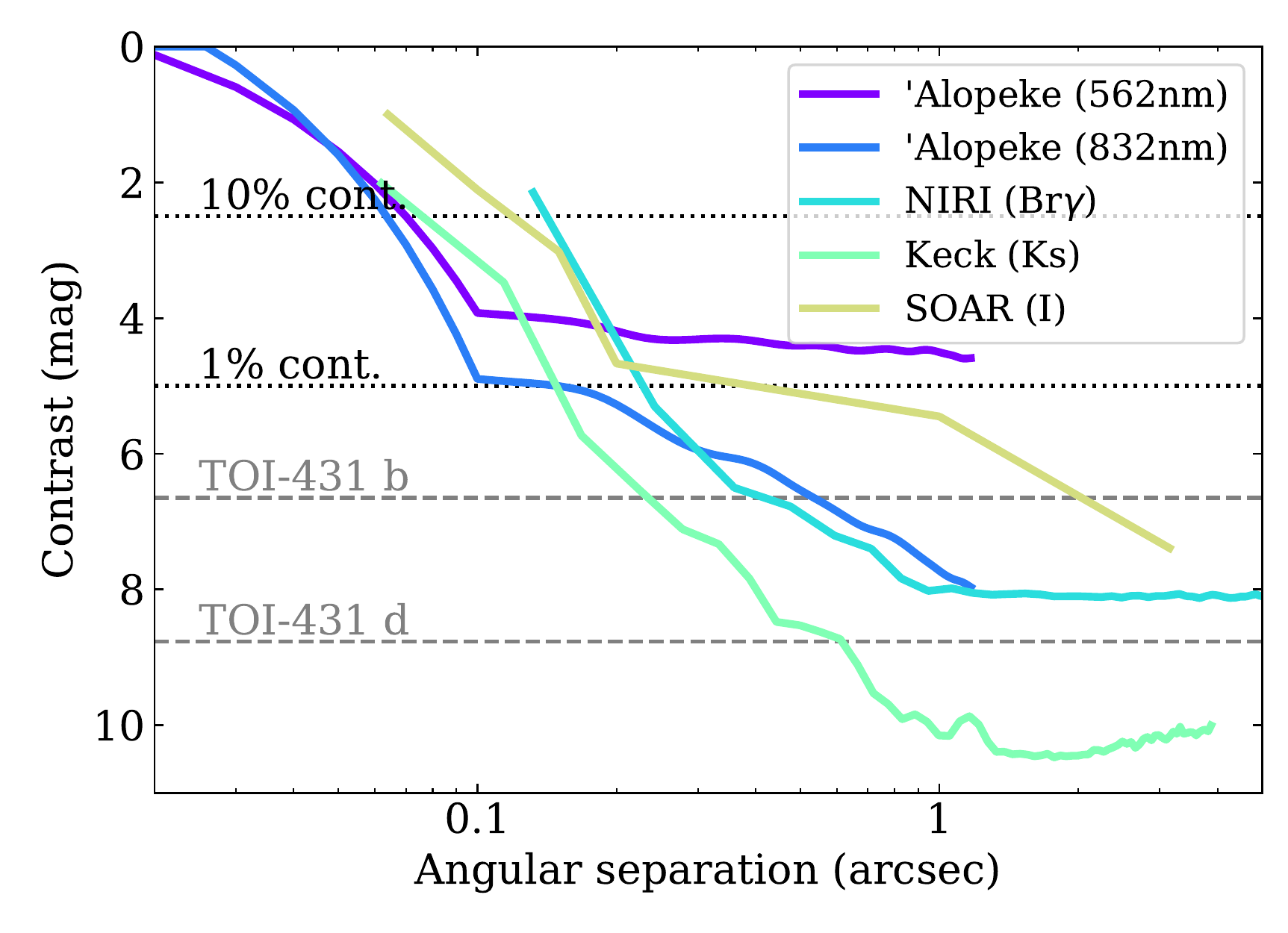}
    \end{subfigure}
    \begin{subfigure}{0.55\textwidth}
    \includegraphics[width=\columnwidth]{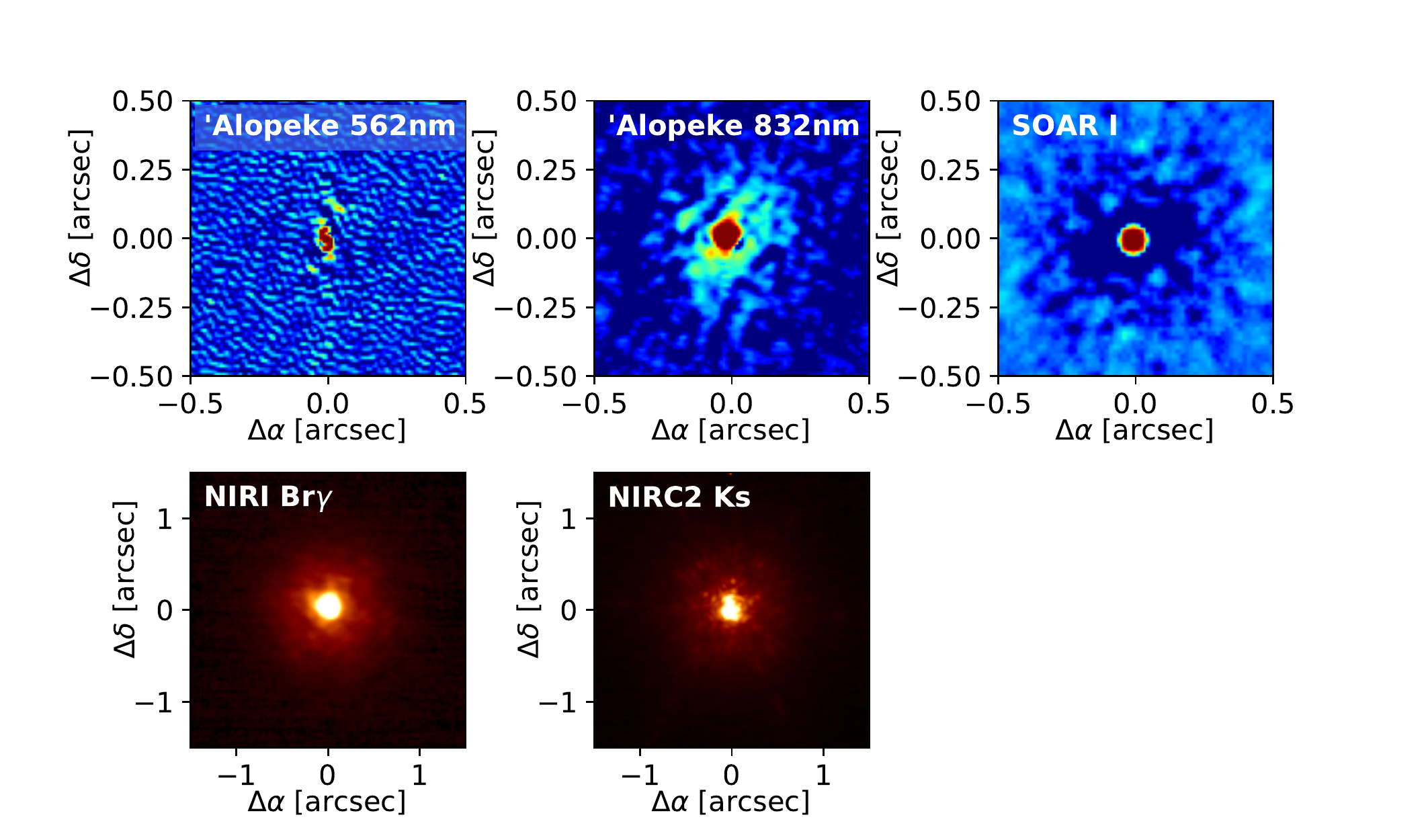}
    \end{subfigure}
    \caption{\textit{Left:} 5$\sigma$ contrast curves for all of the sources of high-resolution imaging described in Section \ref{sec:hri}. The 10 and 1 per cent contamination limits are given as the black dotted lines. The grey dashed lines labelled TOI-431\,b and d represent the maximum contrast magnitude that a blended source could have in order to mimic the planetary transit depth if it were an eclipsing binary. \textit{Right:} a compilation of reconstructed images from 'Alopeke and SOAR and AO images from NIRI and NIRC2, with the instrument and filter labelled. No additional companions are seen.}
    \label{fig:contrastcurves}
\end{figure*}

High angular resolution imaging is needed to search for nearby sources that can contaminate the \TESS\ photometry, resulting in an underestimated planetary radius, or that can be the source of astrophysical false positives, such as background eclipsing binaries. The contrast curves from all of the sources of high resolution imaging described below are displayed in Fig. \ref{fig:contrastcurves}.

\subsubsection{SOAR HRCam} % Carl Ziegler

We searched for stellar companions to TOI-431 with speckle imaging with the 4.1-m Southern Astrophysical Research (SOAR) telescope \citep{Andrei2018} on UT March 17 2019, observing in the Cousins I-band, a similar visible bandpass to TESS. More details of the observation are available in \citet{Ziegler2020}. The 5$\sigma$ detection sensitivity and speckle auto-correlation functions from the observations are shown in Fig. \ref{fig:contrastcurves}. No nearby stars were detected within 3\arcsec of TOI-431 in the SOAR observations.
 
\subsubsection{Gemini NIRI} % Elisabeth Matthews

We collected high resolution adaptive optics observations using the Gemini/NIRI instrument \citep{Hodapp2003} on UT March 18 2019. We collected nine images in the Br$\gamma$ filter, with exposure time 0.6\,s per image. We dithered the telescope by ~2'' between each exposure, allowing for a sky background to be constructed from the science frames themselves. We corrected individual frames for bad pixels, subtracted the sky background, and flat-corrected frames, and then co-added the stack of images with the stellar position aligned. To calculate the sensitivity of these observations, we inject fake companions and measure their S/N, and scale the brightness of these fake companions until they are recovered at 5$\sigma$. This is repeated at a number of locations in the image. We average our sensitivity over position angle, and show the sensitivity as a function of radius in Fig. \ref{fig:contrastcurves}. Our observations are sensitive to companions 4.6\,mag fainter than the host at 0.2'', and 8.1\,mag fainter than the host in the background limited regime, at separations greater than 1''.

\subsubsection{Gemini 'Alopeke} % Steve Howell

TOI-431 was observed on UT Oct 15 2019 using the ‘Alopeke speckle instrument on Gemini-North\footnote{https://www.gemini.edu/sciops/instruments/alopeke-zorro/}. ‘Alopeke provides simultaneous speckle imaging in two bands, 562\,nm and 832\,nm, with output data products including a reconstructed image, and robust limits on companion detections \citep{howell2011}. Fig.~\ref{fig:contrastcurves} shows our results in both 562\,nm and 832\,nm filters. Fig. \ref{fig:contrastcurves} (right) shows the 832\,nm reconstructed speckle image from which we find that TOI-431 is a single star with no companion brighter than within 5-8 magnitudes of TOI-431 detected within 1.2\,\arcsec.  
%[Here you could set limits on spectral type based on the spectral type of TOI-431 minus 5-8 magnitudes).
The inner working angle of the ‘Alopeke observations are 17 mas at 562\,nm and 28 mas at 832\,nm. %These inner limits correspond to spatial limits of xx and yy au respectively at the distance of TOI-431.

\subsubsection{Keck NIRC2}

As part of our standard process for validating transiting exoplanets to assess the possible contamination of bound or unbound companions on the derived planetary radii \citep{Ciardi2015}, we observed TOI-431 with infrared high-resolution Adaptive Optics (AO) imaging at Keck Observatory.  The Keck Observatory observations were made with the NIRC2 instrument on Keck-II behind the natural guide star AO system.  The observations were made on UT March 25 2019 in the standard 3-point dither pattern that is used with NIRC2 to avoid the left lower quadrant of the detector, which is typically noisier than the other three quadrants. The dither pattern step size was $3\arcsec$ and was performed three times.

The observations were made in the $Ks$ filter $(\lambda_o = 2.196; \Delta\lambda = 0.336\mu$m) with an integration time of 1 second and 20 coadds per frame for a total of 300 seconds on target.  The camera was in the narrow-angle mode with a full field of view of $\sim10\arcsec$ and a pixel scale of $0.099442\arcsec$ per pixel. The Keck AO observations revealed no additional stellar companions detected to within a resolution $\sim 0.05\arcsec$ FWHM (Fig. \ref{fig:contrastcurves}).

The sensitivities of the final combined AO image were determined by injecting simulated sources azimuthally around the primary target every $45^\circ $ at separations of integer multiples of the central source's FWHM \citep{Furlan2017}. The brightness of each injected source was scaled until standard aperture photometry detected it with $5\sigma $ significance. The resulting brightness of the injected sources relative to the target set the contrast limits at that injection location. The final $5\sigma $ limit at each separation was determined from the average of all of the determined limits at that separation and the uncertainty on the limit was set by the rms dispersion of the azimuthal slices at a given radial distance. The sensitivity curve is shown in Fig. \ref{fig:contrastcurves} (left), along with an image centred on the primary target showing no other companion stars (right).

\subsubsection{Unbound Blended Source Confidence (BSC) analysis} % Jorge Lillo-Box
\label{sec:BSC}

\begin{figure}
    \centering
    \includegraphics[width=\columnwidth]{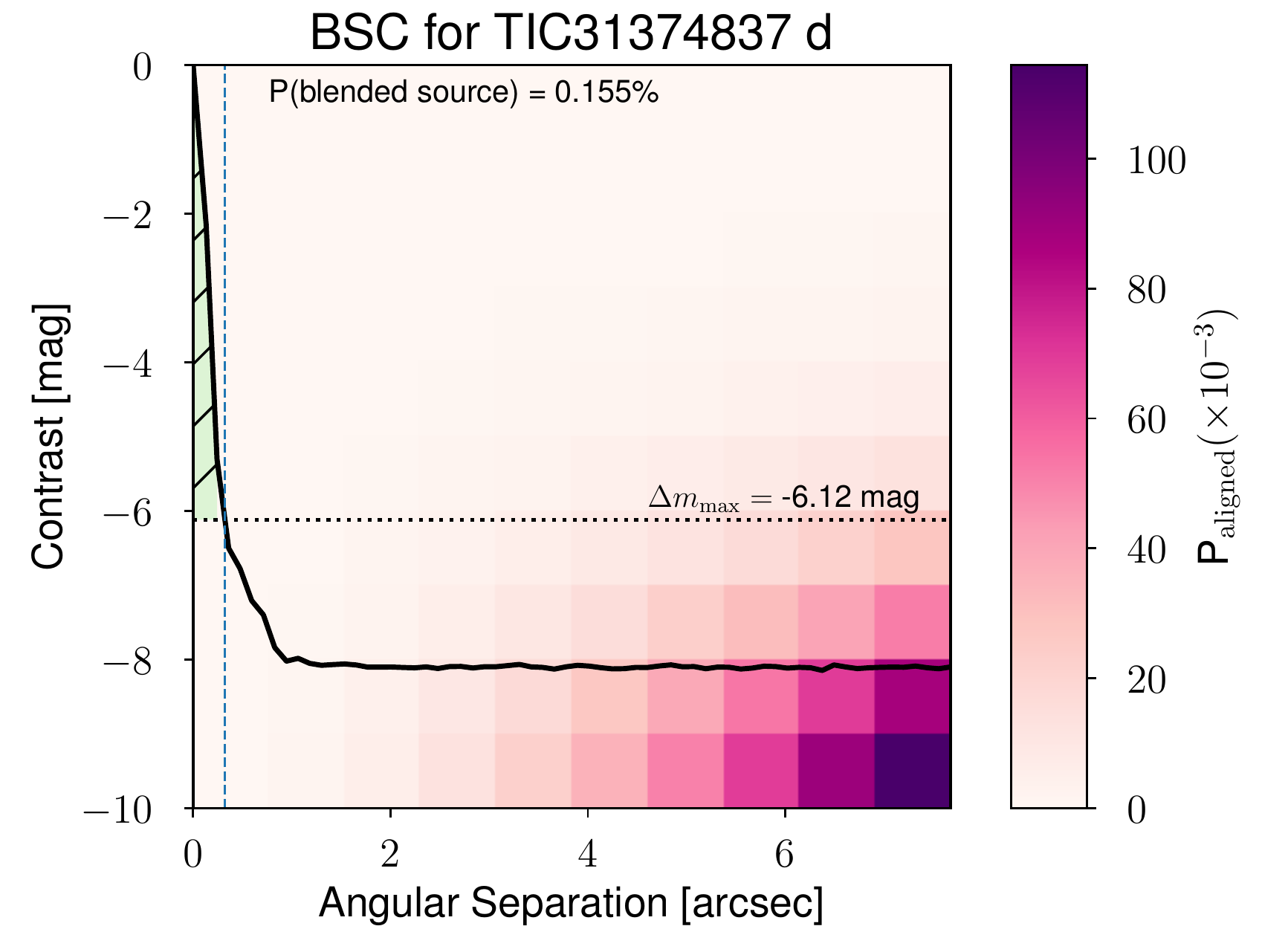}
    \caption{Contrast curve of TOI-431 from the Keck/NIRC2 instrument for the Ks filter (solid black line). The colour ($P_{\rm aligned}$) on each angular separation and contrast bin represents the probability of a chance-aligned source with these properties at the location of the target, based on TRILEGAL model (see Sect.~\ref{sec:BSC} within the main text). The maximum contrast of a blended binary capable of mimicking the planet transit depth is shown as a dotted horizontal line. The hatched green region between the contrast curve and the maximum contrast of a blended binary ($\Delta m_{max}$ line) represents the non-explored regime by the high-spatial resolution image. P(blended source) is the Blended Source Confidence (BSC), and this corresponds to the integration of $P_{\rm aligned}$ over the shaded region.}
    \label{fig:bsc_zorro832}
\end{figure}

We finally analyse all contrast light curves available for this target to estimate the probability of contamination from unbound blended sources in the \TESS\ aperture that are undetectable from the available high-resolution images. This probability is called the Blended Source Confidence (BSC), and the steps for estimating it are fully described in \citet{lillo-box14b}. We use a Python implementation of this approach (\texttt{bsc}, by J. Lillo-Box) which uses the TRILEGAL\footnote{\url{http://stev.oapd.inaf.it/cgi-bin/trilegal}} galactic model \citep[v1.6][]{girardi12} to retrieve a simulated source population of the region around the corresponding target\footnote{This is done in Python by using the \citet{astrobase} implementation.}. This is used to compute the density of stars around the target position (radius $r=1^{\circ}$), and to derive the probability of chance alignment at a given contrast magnitude and separation. We used the default parameters for the bulge, halo, thin/thick disks, and the lognormal initial mass function from \cite{chabrier01}.

The contrast curves of the high-spatial resolution images are used to constrain this parameter space and estimate the final probability of undetected potentially contaminating sources. We consider as potentially contaminating sources those with a maximum contrast magnitude corresponding to $\Delta m_{\rm max} = -2.5\log{\delta}$, with $\delta$ being the transit depth of the candidate planet in the \TESS\ band. This offset from the target star magnitude gives the maximum magnitude that a blended star can have to mimic this transit depth. We convert the depth in the \TESS\ passband to each filter (namely 562\,nm and 832\,nm for the Gemini/'Alopeke images and Ks for the rest) by using simple conversions using the TIC catalogue magnitudes and linking the 562\,nm filter to the SDSSr band, the 832\,nm filter to the SDSSz band and the Ks band to the 2MASS Ks filter. The corresponding conversions imply $\Delta m_{\rm 562\,nm} = 0.954\Delta m_{\rm TESS}$, $\Delta m_{\rm 832\,nm} = 0.920\Delta m_{\rm TESS}$, and $\Delta m_{\rm Ks} = 0.919\Delta m_{\rm TESS}$. In Fig.~\ref{fig:bsc_zorro832} we show an example of the BSC calculation for the Keck/NIRC2 image that illustrates the method. 

We applied this technique to TOI-431. The transits of the two planets in this system could be mimicked by blended eclipsing binaries with magnitude contrasts up to  $\Delta m_{\rm b,max} = 6.65$~mag and $\Delta m_{\rm d,max} = 8.76$~mag in the \TESS\ passband. This analysis is then especially relevant for the smallest planet in the system as the probability of a chance-aligned star increases rapidly with fainter magnitudes. However, the high quality of the high-spatial resolution images provide a very low probability for an undetected source capable of mimicking the transit signal. For TOI-431\,b, we find 0.034 per cent ('Alopeke/562\,nm), 0.019 per cent ('Alopeke/832\,nm), 0.13 per cent (Keck/NIRC2/Ks), and 0.54 per cent (Gemini-North/NIRI/Ks). For TOI-431\,d we find  0.009 per cent ('Alopeke/562\,nm), 0.002 per cent ('Alopeke/832\,nm), 0.04 per cent (Keck/NIRC2/Ks), and 0.16 per cent (Gemini-North/NIRI/Ks).

%%%%%%%%%%%%%%%%%%%%%%%%%%%%%%%%%%%%%%%%%%%%%%%%%%%%%%%%%%%%%%%%%%%%%%%%
\subsection{Stellar analysis}

\begin{center}
\begin{table}
    \caption{Stellar parameters for TOI-431. Section references describing the method used to find the parameters are given in the Table footer.}
    \label{tab:stellarparams}
    \begin{threeparttable}
    \begin{tabularx}{\columnwidth}{ l l X }
    \hline
    \textbf{Parameter (unit)} & \textbf{Value} & \textbf{Ref} \\
    \hline
    Effective temperature $T_{\rm eff}$ (K)                     & 4850 $\pm$ 75         & 1 \\
    Surface gravity $\log g$ (cgs)                              & 4.60 $\pm$ 0.06       & 1 \\
    Microturbulence $V_{\rm tur, mic}$ (km s$^{-1}$)            & 0.8 $\pm$ 0.1 (fixed) & 1 \\
    Macroscopic turbulence $V_{\rm tur, mac}$ (km s$^{-1}$)     & 0.5 $\pm$ 0.1 (fixed) & 1 \\
    Bolometric flux $F_{bol}$ (10$^{-9}$ erg s$^{-1}$ cm$^{-2}$) & 7.98 $\pm$ 0.19       & 2 \\
    Stellar radius $\rstar$ (\rsun)                             & 0.731 $\pm$ 0.022 & 2 \\
    Stellar mass $\mstar$ (\msun)                               & 0.78 $\pm$ 0.07   & 2 \\
    Rotation period $P_{\rm rot}$ (days)                        & 30.5 $\pm$ 0.7        & 3 \\
    \vsini                                                      & 2.5 $\pm$ 0.6         & 1 \\
    \hline
    \textbf{Chemical Abundances (dex)} & \textbf{Value} & \textbf{Ref} \\
    \hline
    Metallicity $[$Fe/H$]$ & 0.2 $\pm$ 0.05 & 1 \\
    $[$NaI/H$]$ & 0.22 $\pm$ 0.14 & 4 \\
    $[$MgI/H$]$ & 0.10 $\pm$ 0.07 & 4 \\
    $[$AlI/H$]$ & 0.21 $\pm$ 0.10 & 4 \\
    $[$SiI/H$]$ & 0.11 $\pm$ 0.13 & 4 \\
    $[$CaI/H$]$ & 0.06 $\pm$ 0.15 & 4 \\
    $[$TiI/H$]$ & 0.17 $\pm$ 0.17 & 4 \\
    $[$CrI/H$]$ & 0.12 $\pm$ 0.11 & 4 \\
    $[$NiI/H$]$ & 0.14 $\pm$ 0.08 & 4 \\
    \hline
    \end{tabularx}
    \begin{tablenotes}
    \item 1: Section \ref{sec:spectroscopic-malcolm}
    \item 2: Section \ref{sec:SED-keivan}
    \item 3: From WASP-South, see Section \ref{sec:stellaractivity}
    \item 4: Section \ref{sec:spectroscopic-vardan}
    \end{tablenotes}
    \end{threeparttable}
\end{table}
\end{center}

The parameters of the host star are required in order to derive precise values for the planetary ages, as well as the masses and radii, leading to bulk densities. This requires a good spectrum with high enough signal-to-noise and high spectral resolution. Our radial velocity spectra fulfil these requirements after co-adding the 124 individual HARPS spectra, resulting in a spectrum with a signal-to-noise of about 380 per pixel at 5950\AA. We perform 2 independent spectroscopic analysis methods to derive the host star parameters, and further SED fitting. 
\bigskip

\subsubsection{Method 1: equivalent widths with ARES+MOOG:} % Vardan Adibekyan and Sergio Sousa
\label{sec:spectroscopic-vardan}

The stellar atmospheric parameters ($T_{\mathrm{eff}}$, $\log g$, microturbulence, and [Fe/H]) and respective error bars were derived using the methodology described in \citet[][]{Sousa-14, Santos-13}. In brief, we make use of the equivalent widths (EW) of iron lines, as measured in the combined HARPS spectrum of TOI-431 using the ARES v2 code\footnote{The last version of ARES code (ARES v2) can be downloaded at http://www.astro.up.pt/$\sim$sousasag/ares.} \citep{Sousa-15}, and we assume ionization and excitation equilibrium. The process makes use of a grid of Kurucz model atmospheres \citep{Kurucz-93} and the radiative transfer code MOOG \citep{Sneden-73}. This analysis results in values of effective temperature $T_{\mathrm{eff}} = 4740 \pm 94$\,K, surface gravity log $g = 4.20 \pm 0.27$, microturbulence V$_{\mathrm{tur}} = 0.62 \pm 0.28$, and metallicity [Fe/H] $= 0.06 \pm 0.04$\,dex. The value for log $g$ can be corrected according to \citet{Mortier2014}, to give $4.46 \pm 0.27$ (corrected for asteroseismology log $g$ values) and $4.63 \pm 0.28$ (corrected for transit log $g$ values).
Stellar abundances of the elements were derived using the classical curve-of-growth analysis method assuming local thermodynamic equilibrium \citep[e.g.][]{Adibekyan-12, Adibekyan-15, Delgado-17}. For the abundance determinations we used the same tools and models as for stellar parameter determination. Unfortunately, due to the low \teff\ of this star, we could not determine reliable abundances of carbon and oxygen. The derived abundances are presented in Table \ref{tab:stellarparams} and they are normal for a star with a metallicity close to solar.

%The derived abundances are [NaI/H] $= 0.22 \pm 0.14$, [MgI/H] $= 0.10 \pm 0.07$, [AlI/H] $= 0.21 \pm 0.10$, [SiI/H] $= 0.11 \pm 0.13$, [CaI/H] $= 0.06 \pm 0.15$, [TiI/H] $= 0.17 \pm 0.17$, [CrI/H] $= 0.12 \pm 0.11$, and [NiI/H] $= 0.14 \pm 0.08$, and these are also presented in Table \ref{tab:stellarparams}. 

In addition, we derived an estimated age by using the ratios of certain elements (the so-called chemical clocks) and the formulas presented in \citet{Delgado-19}. Since this star has a close to solar metallicity and is very cool (and thus probably outside the applicability limits of formulas using stellar parameters in addition to the chemical clock) we chose to use the 1D formulas presented in Table 5 of \citet{Delgado-19}. Due to the high error in Sr abundances we derived ages only from the abundance ratios [Y/Mg], [Y/Zn], [Y/Ti], [Y/Si], [Mg/Fe], [Ti/Fe], [Si/Fe] and [Zn/Fe]. The abundance errors of cool stars are quite large and in turn the individual age errors of each chemical clock are also large ($\gtrsim$\,3\,Gyr) but the dispersion among them is smaller. We obtained a weighted average age of 5.1\,$\pm$\,0.6 Gyr which is significantly older than the age obtained in Section \ref{sec:SED-keivan}. Nevertheless, we note that ages for very cool stars obtained from chemical clocks are affected by large errors and must be taken with caution.

\subsubsection{Method 2: synthesis of the entire optical spectrum} % Malcolm Fridlund
\label{sec:spectroscopic-malcolm}

We also derived stellar properties by analysing parts of the optical spectrum in a different way by comparing the normalized, co-added spectrum with modelled synthetic spectra obtained with the Spectroscopy Made Easy (SME) package \citep{1996A&AS..118..595V,2017A&A...597A..16P} version 5.22, with atomic parameters from the VALD database \citep{1995A&AS..112..525P}. The 1-D, plane-parallel LTE synthetic spectra are calculated using  stellar parameters obtained from either photometry or a visual inspection of the spectrum as a starting point. The synthetic spectrum is automatically then compared to a grid of stellar atmospheric models. The grid we used in this case is based on the MARCS models \citep{2008A&A...486..951G}. An iterative $\chi^2$ minimization procedure is followed until no improvement is achieved. We refer to recent papers, e.g., \citet{2018A&A...618A..33P} and \citet{2008A&A...486..951G} for details about the method. In order to limit the number of free parameters we used empirical calibrations for the \vmic\ and \vmac\ turbulence velocities \citep{2010MNRAS.405.1907B, 2014MNRAS.444.3592D}. The value of \teff\ was determined from fitting the Balmer \halpha~line wings. We used the derived \teff\ to fit a large sample of \fei, \mgi and \cai lines, all with well established atomic parameters in order to derive the abundance, \feh, the rotation, and the surface gravity, \logg. We found the star to be slowly rotating, with \vsini~=\,2.5~$\pm$\, 0.6 \kms.  The star is cool, and the effective temperature as derived from the \halpha~line wings is \teff\,=\,$4846 \pm 73$~K. Using this value for \teff\ we found the \feh~to be $0.20\,\pm\,0.05$ and the surface gravity \logg~to be $4.60\,\pm\,0.06$ (Table~\ref{tab:stellarparams}).

In order to check our result, we also analysed the same co-added spectrum using the public software package SpecMatch-Emp \citep{2017ApJ...836...77Y}. This program extracts part of the spectrum and attempts to match it to a library of about 400 well characterized spectra of all types. Our input spectrum has to conform to the format of SpecMatch-Emp and we refer to \citet{2018AJ....155..127H} to describe our procedure for doing this. We derive a \teff\ of $4776 \pm 110$~K, an iron abundance of \feh = $0.15 \pm 0.09$ dex, and a stellar radius of $R_\star\,=\,0.76\,\pm\,0.18$~\rsun. The former two values are in good agreement with the results from the SME analysis. 

Because of the higher precision in the SME analysis, the final adopted value of \teff\ for TOI-431 is \mbox{$4850\,\pm\,75$~K}. Note that the error here is the internal errors in the synthesis of the spectra and does not include the inherent errors of the model grid itself, as well as those errors caused by using 1-D models. 

The results from this method are in agreement with those found in Section \ref{sec:spectroscopic-vardan}, with $T_{\rm{eff}}$ and [Fe/H] (using SpecMatch-Emp) agreeing within error. The value for log$g$ also agrees with the corrected log$g$ values from the previous method. We therefore adopt the results from this method to take forward.

\subsubsection{SED fitting} % Keivan Stassun
\label{sec:SED-keivan}

\begin{figure}
    \centering
    \includegraphics[width=\columnwidth]{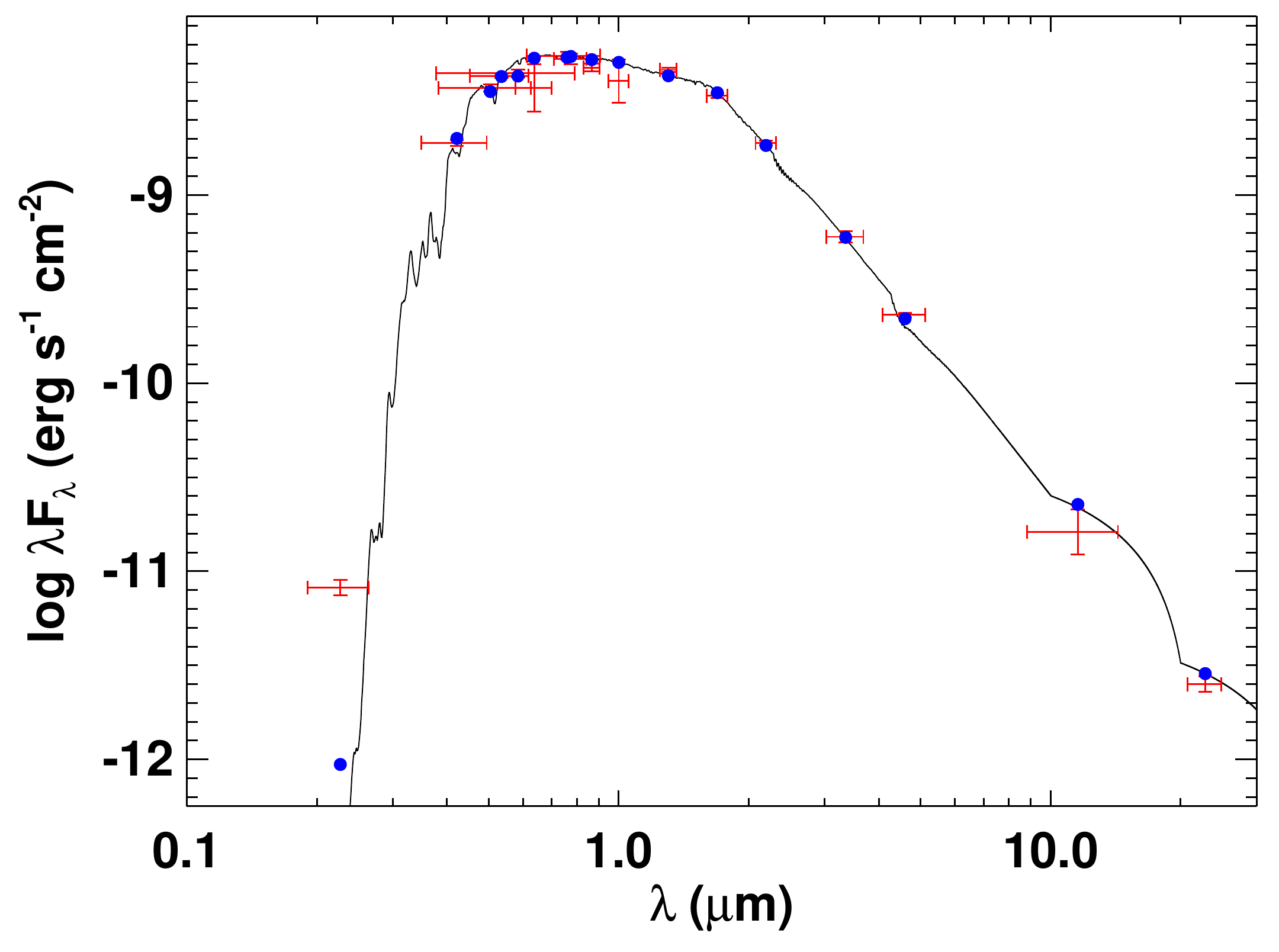}
    \caption{Spectral energy distribution (SED) of TOI-431. Red symbols represent the observed photometric measurements, where the horizontal bars represent the effective width of the passband. Blue symbols are the model fluxes from the best-fit Kurucz atmosphere model (black).}
    \label{fig:sed}
\end{figure}

As an independent check on the derived stellar parameters, and in order to determine an estimate for stellar age, we performed an analysis of the broadband spectral energy distribution (SED). Together with the {\it Gaia\/} EDR3 parallax, we determine an empirical measurement of the stellar radius following the procedures described in \citet{Stassun2016,Stassun2017,Stassun2018}. We pulled the $B_T V_T$ magnitudes from {\it Tycho-2}, the $grizy$ magnitudes from Pan-STARRS, the $JHK_S$ magnitudes from {\it 2MASS}, the W1--W4 magnitudes from {\it WISE}, and the $G G_{\rm RP} G_{\rm BP}$ magnitudes from {\it Gaia}. Together, the available photometry spans the full stellar SED over the wavelength range 0.35--22~$\mu$m (see Figure~\ref{fig:sed}). In addition, we pulled the NUV flux from {\it GALEX} in order to assess the level of chromospheric activity, if any. 

We performed a fit using Kurucz stellar atmosphere models, with the effective temperature ($T_{\rm eff}$) and metallicity ([Fe/H]) adopted from the spectroscopic analysis (Section \ref{sec:spectroscopic-malcolm}). The extinction ($A_V$) was set to zero because of the star being very nearby (Table \ref{tab:toi-431}). The resulting fit is excellent (Figure~\ref{fig:sed}) with a reduced $\chi^2$ of 3.3 (excluding the {\it GALEX} NUV flux, which is consistent with a modest level of chromospheric activity; see below). Integrating the (unreddened) model SED gives the bolometric flux at Earth of $F_{\rm bol} =  7.98 \pm 0.19 \times 10^{-9}$ erg~s$^{-1}$~cm$^{-2}$. Taking the $F_{\rm bol}$ and $T_{\rm eff}$ together with the {\it Gaia\/} EDR3 parallax, with no systematic offset applied \citep[see, e.g.,][]{StassunTorres:2021}, gives the stellar radius as $R =  0.731 \pm 0.022$~R$_\odot$. Finally, estimating the stellar mass from the empirical relations of \citet{Torres2010} and a 6\% error from the empirical relation itself gives $M = 0.77 \pm 0.05 M_\odot$, whereas the mass estimated empirically from the stellar radius together with the spectroscopic $\log g$ gives $M = 0.78 \pm 0.07 M_\odot$. 

We can also estimate the stellar age by taking advantage of the observed chromospheric activity together with empirical age-activity-rotation relations. For example, taking the chromospheric activity indicator $\log R'_{HK} = -4.69 \pm 0.05$ from the archival HARPS data and applying the empirical relations of \citet{Mamajek2008} gives a predicted age of $1.9 \pm 0.3$~Gyr. Finally, we can further corroborate the activity-based age estimate by also using empirical relations to predict the stellar rotation period from the activity. For example, the empirical relation between $R'_{HK}$ and rotation period from \citet{Mamajek2008} predicts a rotation period for this star of $29.8 \pm 3.7$~d, which is compatible with the rotation period inferred from the WASP-South observations (see Section \ref{sec:stellaractivity}). All of the stellar parameter values derived in this section can also be found in Table \ref{tab:stellarparams}. 

%which is compatible with the rotation period inferred from the spectroscopic $v\sin i$ which gives $P_{\rm rot}/\sin i = 19.2 \pm 4.8$~d. 

%%%%%%%%%%%%%%%%%%%%%%%%%%%%%%%%%%%%%%%%%%%%%%%%%%%%%%%%%%%%%%%%%%%%%%%%

\subsection{Stellar activity monitoring} % Coel Hellier, Ed Bryant, me
\label{sec:stellaractivity}

\begin{figure}
\includegraphics[width=8cm]{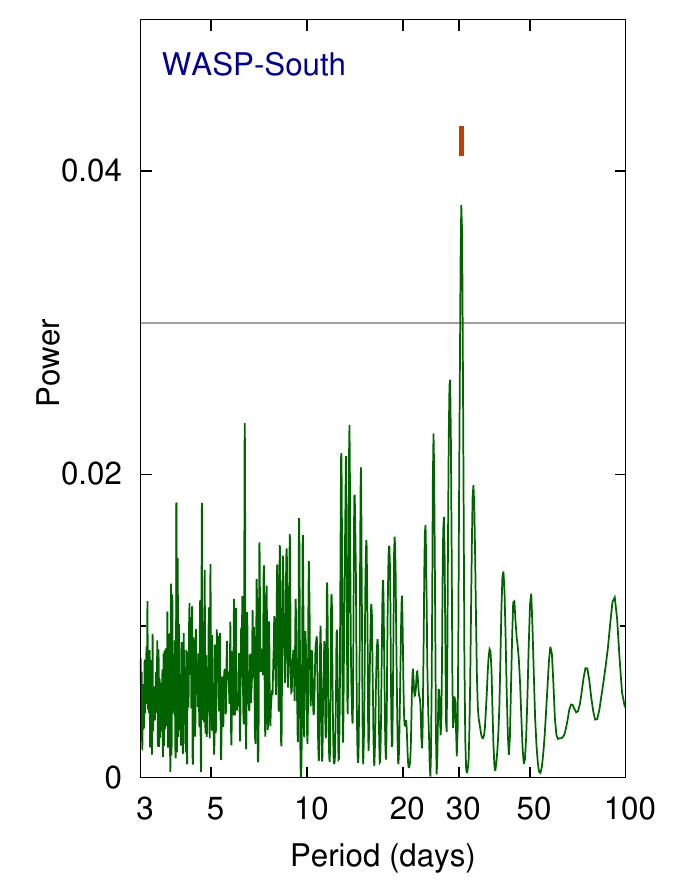}
  \caption{The periodogram of the WASP-South data for TOI-431 from 2012--2014. The orange tick is at 30.5\,d, while the horizontal line is at the estimated 1 per cent false-alarm probability.}
\label{fig:wasp}
\end{figure}

Two instruments were used during different time periods to monitor TOI-431 in order to investigate the rotation period of the star. This is important to disentangle the effect of stellar activity when fitting for any planets present in the system. 

WASP-South, located in Sutherland, South Africa, was the southern station of the WASP transit survey \citep{2006PASP..118.1407P}.  The data reported here were obtained while WASP-South was operating as an array of 85mm, f/1.2 lenses backed by 2048x2048 CCDs, giving a plate scale of $32\arcsec$/pixel.  The observations spanned 180 days in 2012, 175 days in 2013 and 130 days in 2014.  Observations on clear nights, with a typical 10-min cadence, accumulated 52\,800 photometric data points.

We searched the datasets for rotational modulations, both separately and by combining the three years, using the methods described by \citet{2011PASP..123..547M}.  We detect a persistent modulation with an amplitude of 3 mmag and a period of 30.5 $\pm$ 0.7 d (where the error makes allowance for phase shifts caused by changing starspot patterns). The periodogram from the combined 2012--2014 data is shown in Fig.~\ref{fig:wasp}.  The modulation is significant at the 99.9\%\ level (estimated using methods from \citealt{2011PASP..123..547M}).  In principle, it could be caused by any star in the 112$\arcsec$ photometric extraction aperture, but all the other stars are more than 4 magnitudes fainter. 

Given the near-30-day timescale, we need to consider the possibility of contamination by moonlight.  To check this, we made identical analyses of the light curves of 5 other stars of similar brightness nearby in the same field. None of these show the 30.5\,d periodicity. 

A single NGTS telescope was used to monitor TOI-431 between the dates of 2019 October 11 and 2020 January 20. During this time period a total of 79011 images were taken with an exposure time of 10\,seconds using the custom NGTS filter (520 - 890\,nm). This data shows a significant periodicity at 15.5 days, at approximately half the period of the WASP-South modulation.

As the WASP-South period agrees with the activity signal we see in the HARPS data (see Fig. \ref{fig:RVlombscargle}), we therefore take the 30.5\,d period value forward.

\section{The Joint Fit} \label{sec:jointfit}
\begin{figure*}
    \begin{subfigure}{\textwidth}
        \centering
        \includegraphics[width=\textwidth]{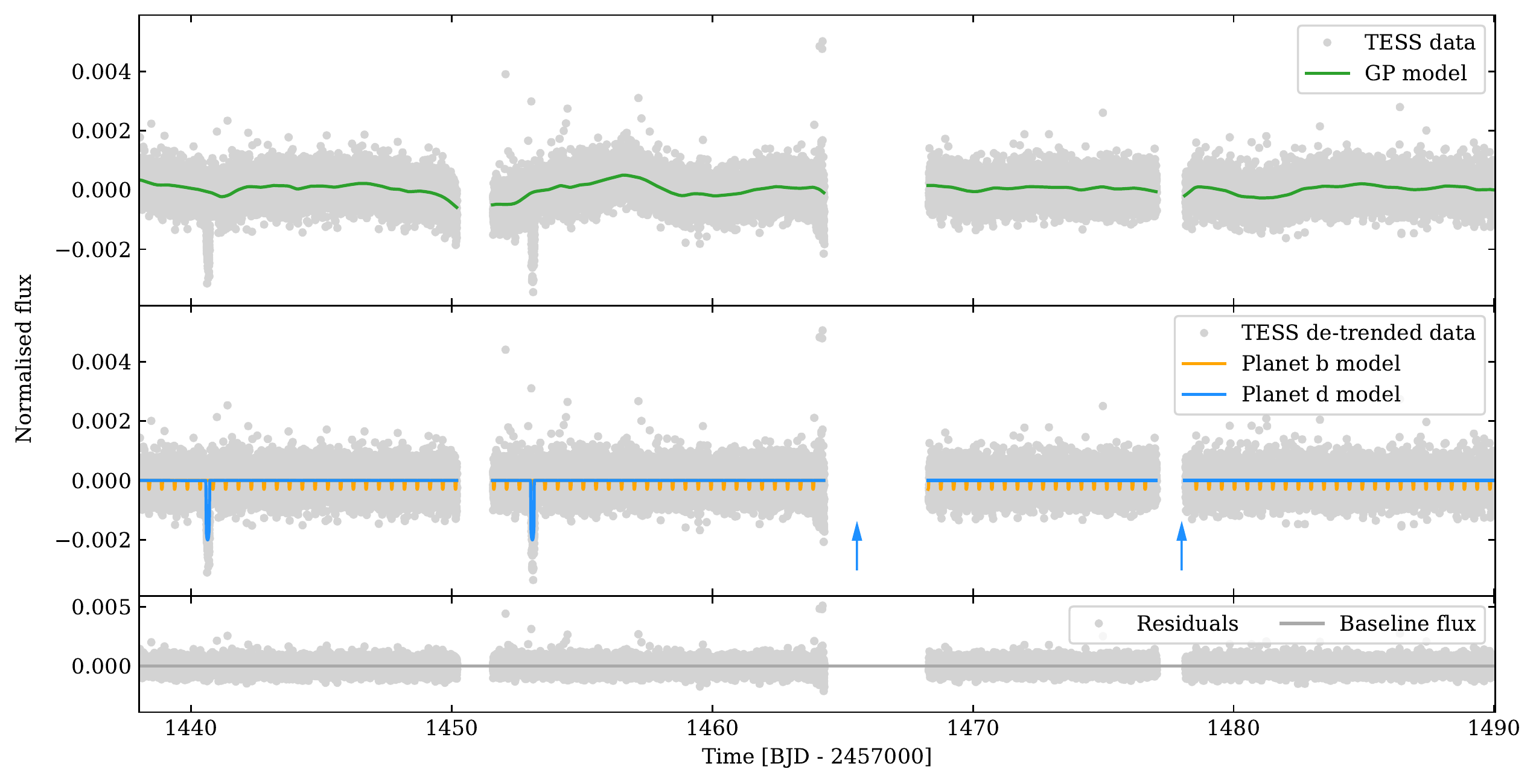}
    \end{subfigure}
    \newline
    \begin{subfigure}{\textwidth}
        \centering
        \includegraphics[width=\textwidth]{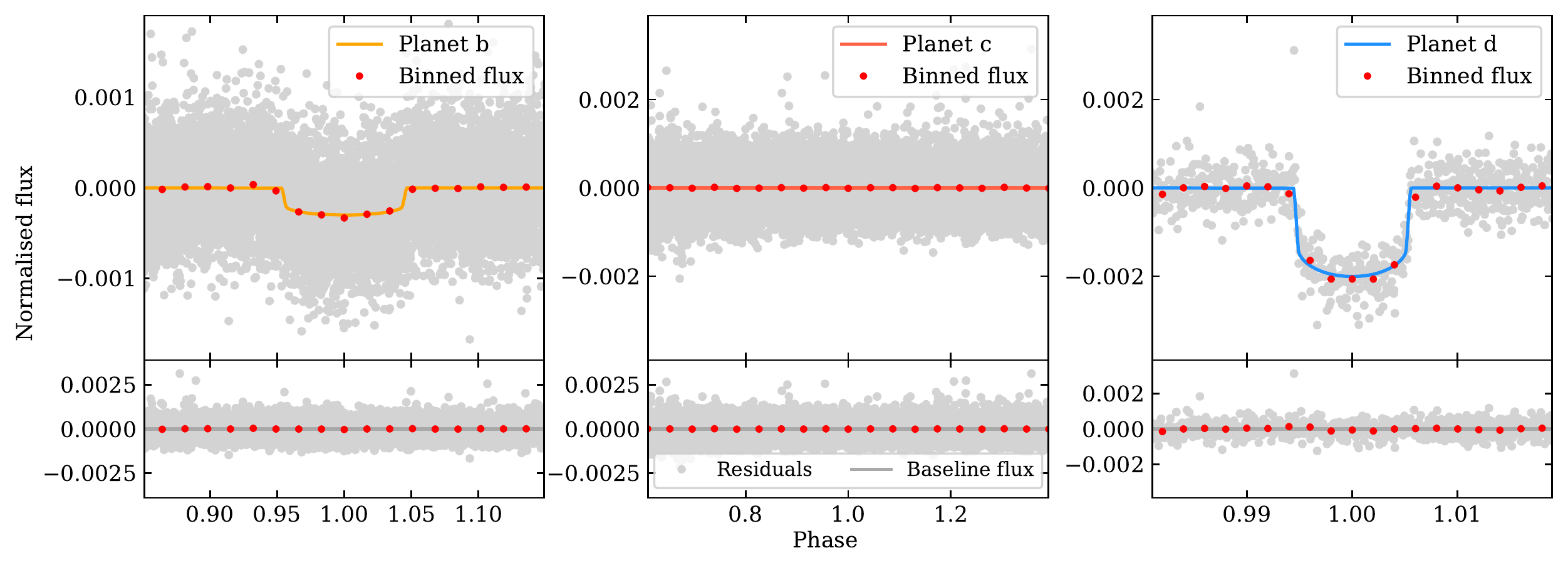}
    \end{subfigure}
    \caption{The \TESS\ data for TOI-431 in Sectors 5 and 6. \textbf{Top plot:} detrending the \TESS\ light curves and fitting models for TOI-431\,b and c. \textit{Top:} the full, 2-min cadence PDCSAP lightcurve, with no detrending for stellar activity, is shown in grey. Each sector has 2 segments of continuous viewing, and the gaps in the data correspond to the spacecraft down-linking the data to Earth after a \TESS\ orbit of 13.7 days. Overlaid in green is the GP model  that has been fit to this data (described in Section \ref{sec:fitphotom}), in order to detrend the stellar activity . \textit{Middle:} the flux detrended with the GP model, with the transit models for TOI-431\,b (orange) and d (blue) overlaid. The expected transit times for the 2 further transits of TOI-431\,d, both of which fall in the data down-link, are marked with blue arrows. \textit{Bottom:} residuals when the best fit model and GP have been subtracted from the PDCSAP flux. The baseline flux (normalised to 0) is shown in dark grey. \textbf{Bottom plot:} phase folds of the \TESS\ data for TOI-431\,b (left), c (middle, with no transit evident), and d (right), with the flux binned as red circles, and the residuals of the folds once the best fit models have been subtracted from the flux shown in the bottom panels.}
\label{fig:TESS}
\end{figure*}

\subsection{The third planet found in the HARPS data}

\begin{figure*}
    \centering
    \includegraphics[width=0.85\textwidth]{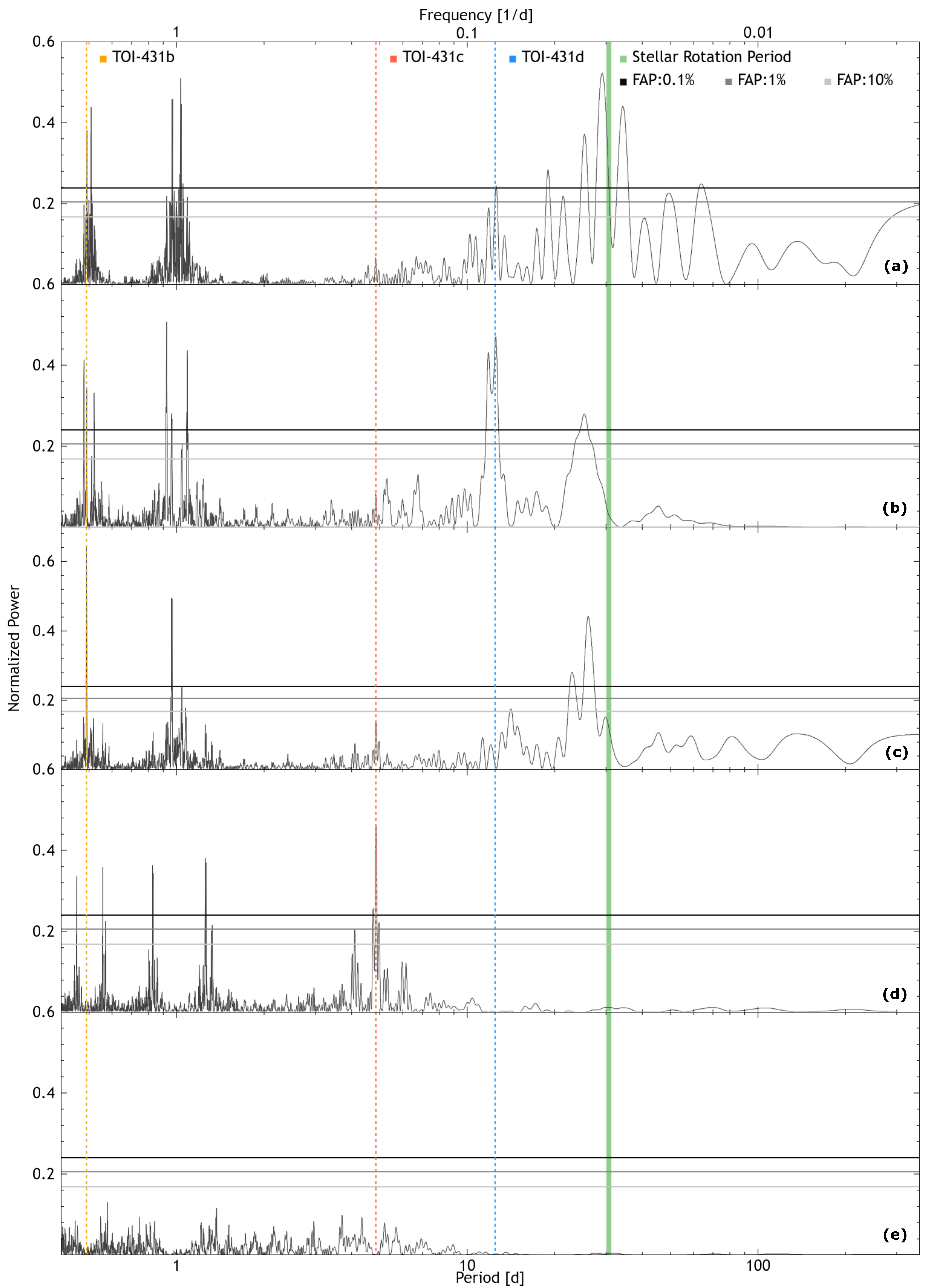}
    \caption{Periodograms for the HARPS data, where (going from top to bottom) the highest power peak has been sequentially removed until there is no power left. The best fit periods (see Table \ref{tab:planetparams}) of TOI-431\,b (yellow), c (red), and d (blue), have been denoted by dotted lines, and the 1 standard deviation interval of the rotation period of the star has been shaded in green. The periodogram for the raw RV data is shown in panel (a); (b) has the stellar activity GP model removed; (c) has the best fit model for planet d removed also. Panel (d) has planet b removed, meaning that there should be no further power left. However, there is a peak evident at 4.85\,days above the 0.1 per cent FAP that does not correlate with any stellar activity indicators, and it is not an alias of any other peaks. Taking this as an extra planet in the system (TOI-431\,c) and removing the best fit model for this leaves a periodogram with no further signals, shown in panel (e).}
    \label{fig:RVlombscargle}
\end{figure*}

We initially ran a joint fit which included only the planets flagged by the \TESS\ pipelines, i.e. TOI-431\,b and d. We then removed the signals of these planets from the raw \HARPS\ radial velocities, and examined the residuals. This led to the discovery of an independent sinusoidal signal being seen as a significant peak in a periodogram of the residuals. This is shown in Fig. \ref{fig:RVlombscargle}: from the periodogram of the raw RV data produced on DACE\footnote{The DACE platform is available at \url{https://dace.unige.ch}}, signals from TOI-431\,b and d can be seen at 0.491 and 12.57\,d respectively, with false-alarm probabilities (FAP) of $< 0.1$ per cent. A large signal can also be seen at 29.06\,d; this is near the rotation period of the star found with WASP-South (see Section \ref{sec:stellaractivity}). Removing the fit for these two planets and the stellar activity reveals another signal at 4.85\,d which does not correlate with any of the activity indicators (FWHM, BIS, S-Index and H$\alpha$-Index; see Fig. \ref{fig:activityindicators} for periodograms of these indicators for both the current and archival HARPS data), and which is not an alias of the other planetary signals. 

Phase folding the \TESS\ photometry on the RV period reveals no transit (see Fig. \ref{fig:TESS}, bottom plot, middle panel). We also attempted to use Transit Least Squares \citep[TLS,][]{Hippke2019} to recover this planet; it did not return any evidence of a transit at or near the RV period. As this planet is not evident in the \TESS\ data, but is large enough to be detectable (see Section \ref{sec:fitresults}), we therefore make the assumption that it does not transit. As such, we conclude that this is a further, apparently non-transiting planet, and include it in the final joint fit model (described in Section \ref{sec:jointfitmodel}) when fitting the RV data.

\subsection{Construction of the joint fit model}
\label{sec:jointfitmodel}

Using the \textsc{exoplanet} package \citep{exoplanet:exoplanet}, we fit the photometry from \TESS, LCOGT, \NGTS, and \Spitzer\ and the RVs from \HARPS\ and HIRES simultaneously with Gaussian Processes (GPs) to remove the effects of stellar variability. \textsc{exoplanet} utilises the light curve modelling package \textsc{Starry} \citep{exoplanet:luger18}, \textsc{PyMC3} \citep{exoplanet:pymc3}, and \textsc{celerite} \citep{exoplanet:celerite} to incorporate GPs. While we use a GP kernel included in the \textsc{exoplanet} package for the \TESS\ data, we construct our own GP kernel using \textsc{PyMC3} for the \HARPS\ and HIRES data. For consistency, all timestamps were converted to the same time system used by \TESS, i.e. BJD\,-\,2457000. All prior distributions set on the fit parameters of this model are given in Table \ref{tab:furtherfitparams}.

\subsubsection{Photometry}
\label{sec:fitphotom}

The flux is normalised to zero for all of the photometry by dividing the individual light curves by the median of their out-of-transit points and taking away one. To model the planetary transits, we used a limb-darkened transit model following the \citet{Kipping2013} quadratic limb-darkening parameterisation, and Keplerian orbit models. This Keplerian orbit model is parameterised for each planet individually in terms of the stellar radius $R_*$ in solar radii, the stellar mass $M_*$ in solar masses, the orbital period $P$ in days, the time of a reference transit $t_0$, the impact parameter $b$, the eccentricity $e$, and the argument of periastron $\omega$. While a similar Keplerian orbit model is parameterised for the third planet, $b$ is not defined in this case as no transit is seen in the photometric data. We find the eccentricity of all planets to be consistent with 0: when eccentricity is a fit parameter in an earlier run of this model, we find the 95 per cent confidence intervals for the eccentricity of TOI-431\,b, c and d to be 0 to 0.28, 0 to 0.22, and 0 to 0.31 respectively.
%$0.07^{+0.21}_{-0.07}$, $0.06^{+0.16}_{-0.06}$ and $0.08^{+0.23}_{-0.08}$, respectively. 
Therefore, we fix $e$ and $\omega$ to 0 for all planets in the final joint fit model. These parameters are then input into light curve models created with \textsc{Starry}, together with parameters for the planetary radii $R_p$, the time series of the data $t$, and the exposure time $t_{\mathrm{exp}}$ of the instrument. As we are modelling multiple planets and multiple instruments with different $t_{\mathrm{exp}}$, a separate light curve model is thus created per instrument for the planets that are expected to have a transit event during that data set. In some cases, TOI-431\,b and d will have model light curves (e.g. in the \TESS\ and \Spitzer\ observations); in others (e.g. the LCOGT and NGTS observations), only TOI-431\,d is expected to be transiting. TOI-431\, c is not seen to transit, therefore we do not need to model it in this way. We use values from the TESS pipelines to inform our priors on the epochs, periods, transit depths and radii of the transiting planets.

\bigskip
\textbf{\textit{TESS}}

Both TOI-431\,b and d are transiting in the \TESS\ light curve, so we first create model light curves for each using \textsc{Starry}. 

As seen in Fig. \ref{fig:TESS}, the TESS Sector 5 and 6 light curves show some stellar variability. This variability was thus modelled with the SHOTerm GP given in \textsc{exoplanet} \footnote{\url{https://docs.exoplanet.codes/en/stable/user/api/\#exoplanet.gp.terms.SHOTerm}}, which represents a stochastically-driven, damped harmonic oscillator. We set this up using the hyperparameters $\log{(s2)}$, $\log{(Sw4)}$, $\log{(w0)}$, and $Q$. The prior on $Q$ was set to $1/\sqrt{2}$. Priors on $\log{(s2)}$ and $\log{(Sw4)}$ were set as normal distributions with a mean equal to the log of the variance of the flux and a standard deviation of 0.1. The prior on $\log{(w0)}$ was also set as a normal distribution but with a mean of 0 and a standard deviation of 0.1 (see Table \ref{tab:furtherfitparams}). 

We then take the sum of our model light curves and subtract these from the total PDCSAP flux, and this resultant transit-free light curve is the data that the GP is trained on to remove the stellar variability. The GP model can be seen in Fig. \ref{fig:TESS} (top plot, top panel), and the resultant best fit model in the middle panel. Further to this, phase folds of the \TESS\ data for all planets in the system can also be seen in Fig. \ref{fig:TESS} (bottom plot), where TOI-431\,c has been folded on its period determined from the radial velocity data, and no dip indicative of a transit can be seen. 

\bigskip
\textbf{LCOGT}

\begin{figure}
    \centering
    \includegraphics[width=\columnwidth]{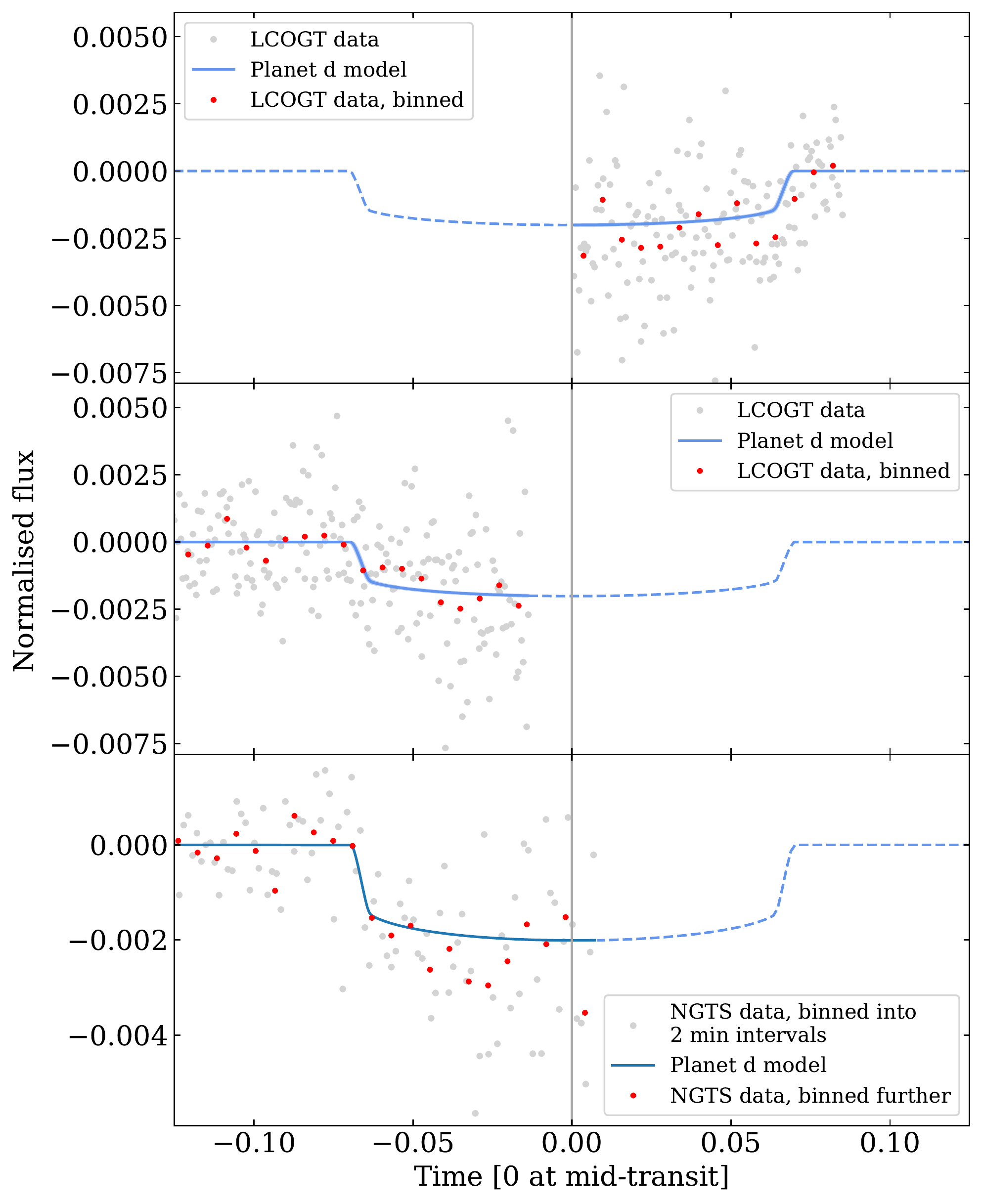}
    \caption{Best fit models of TOI-431\,d to the LCOGT ingress (top), egress (middle) and NGTS light curves (bottom). In the LCOGT panels (top and middle), the observed flux is shown as light grey circles, the binned flux as red circles. In the NGTS panel (bottom), the flux is binned to 2 minute intervals in light grey. In all panels, the fit model is given as the blue line, solid where there are photometry points and dashed where there are not.}
    \label{fig:LCO-NGTS}
\end{figure}

No further detrending to that outlined in Section \ref{LCOphotom} was included for the LCOGT data. Only TOI-431\,d is transiting in this data, so we create a model light curve of TOI-431\,d using \textsc{Starry} (as outlined above) per LCOGT dataset to produce 2 model light curves overall, as there are 2 transit events - an ingress and an egress - on separate nights. For each dataset, we use a normal prior with the model light curve as the mean and a standard deviation set to the error on the LCOGT data points, and this is then compared to the observed light curve. The best fit model for both the ingress and egress data is shown in Fig. \ref{fig:LCO-NGTS} (top 2 panels). 

\bigskip
\textbf{NGTS}

No further detrending was needed for the NGTS data after the pipeline reduction outlined in Section \ref{NGTSphotom}, and again, only TOI-431\,d is evident in this data. Thus the same simple method used for the LCOGT data above is also applied here, creating a singular model light curve of TOI-431\,d for the NGTS data and comparing this to the observed light curve, with a standard deviation set to the error on the NGTS data points. The best fit model for the NGTS data is shown in Fig. \ref{fig:LCO-NGTS} (bottom panel). 

\bigskip
\textbf{\textit{Spitzer} and Pixel Level Decorrelation}

\begin{figure}
    \centering
    \includegraphics[width=\columnwidth]{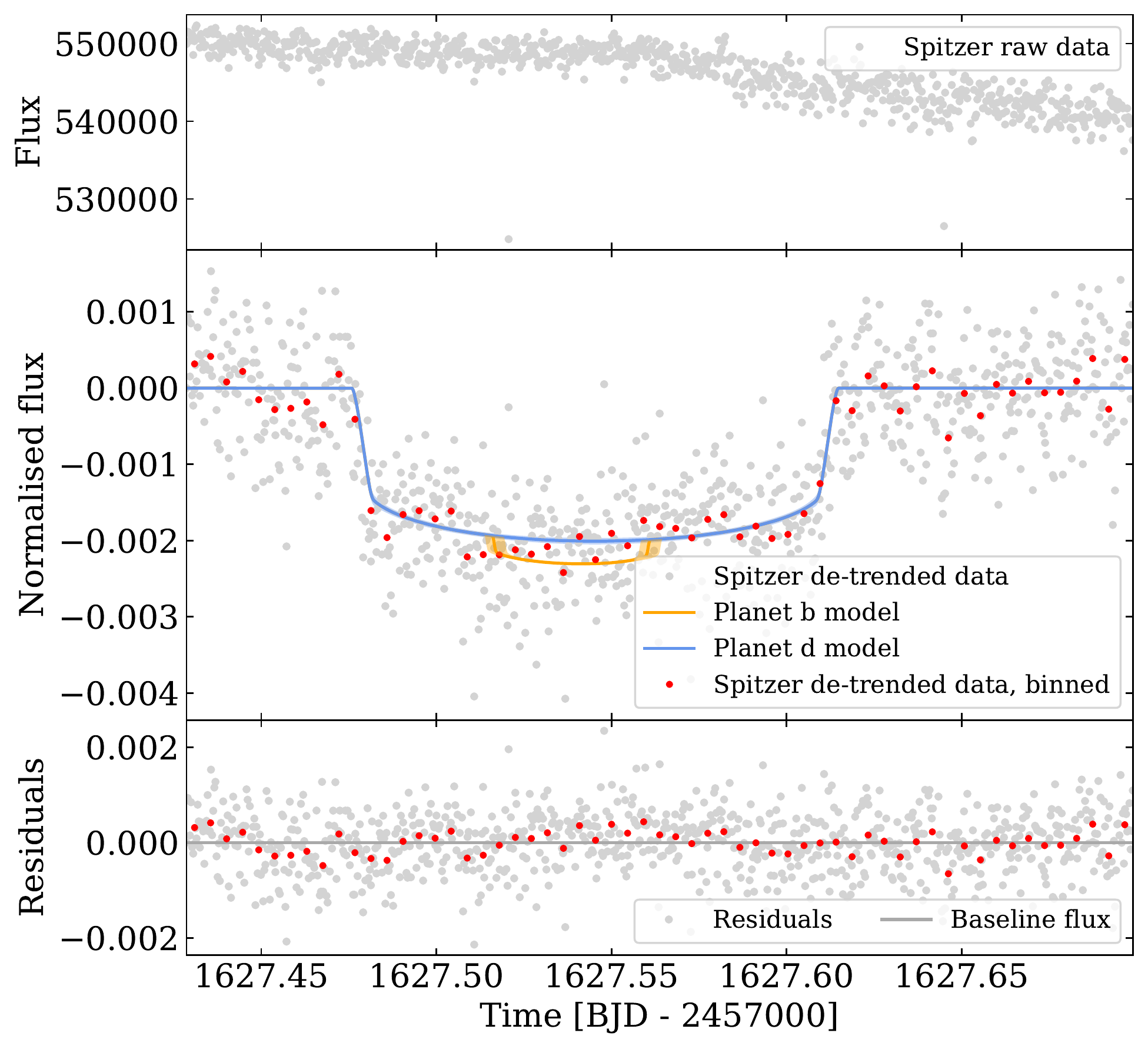}
    \caption{The \Spitzer\ double-transit. \textit{Top:} the raw \Spitzer\ data, without any PLD applied. \textit{Middle:} the \Spitzer\ light curve detrended with PLD in grey and binned as red circles, with the best fit models of planet\,b (orange) and d (blue) overlaid. \textit{Bottom:} the residuals when the best fit model has been subtracted from the detrended flux.}
    \label{fig:spitzer}
\end{figure}

For the \Spitzer\ double-transit observation, model light curves are created for both TOI-431\,b and d. \Spitzer\ data is given as $N$ pixel values on a grid; in this instance, the grid is 3x3 pixels as in figure 1 of \citet{Deming2015}. We follow the Pixel Level Decorrelation (PLD) method of \citet{Deming2015} (summarised below) to remove the systematic effect caused by intra-pixel sensitivity variations. Together with pointing jitter, these variations mask the eclipses of exoplanets in the photometry with intensity fluctuations that must be removed. We outline our PLD implementation as follows:

First, the intensity of pixel $i$ at each time step $t$, i.e. $P_i^t$, is normalised such that the sum of the 9 pixels at one time step is unity, thus removing any astrophysical variations: 
\begin{equation}
    \hat{P}_i^t = \frac{P_i^t}{\sum^N_{i=1} P_i^t}.
\end{equation}

PLD makes the simplification that the total flux observed can be expressed as a linear equation:
\begin{equation}
    \Delta S^t = \sum_{i}^{N} c_i \hat{P}_i^t + DE(t) + ft + gt^2 + h,
\label{equ:PLD}
\end{equation}

where $\Delta S^t$ is the total fluctuation from all sources. The normalised pixel intensities are multiplied by some coefficient $c_i$, and summed with the eclipse model $DE(t)$, a quadratic function of time $ft + gt^2$ which represents the time-dependent ``ramp'', and an offset constant $h$. We use the eclipse model set up earlier using \textsc{exoplanet} as $DE(t)$, where $D$ is the eclipse depth.
This allows us to remove the intra-pixel effect, while solving for the eclipse amplitude and temporal baseline effects. Overall, the PLD  alone has 14 free parameters that we solve for: 9 pixel coefficients, the depth of eclipse and the eclipse model, 2 time coefficients, and an offset term.

We add an additional fit parameter by introducing a \Spitzer\ ``jitter'' term. We can estimate a prior for this fit parameter by removing our best fit model from the total raw flux from Spitzer, and calculating the standard deviation of the residual flux, which is approximately 337\,ppm. 

Our overall model for the \Spitzer\ data is the PLD terms multiplied by the sum of the individual light curve models for each planet, b and d. We use a normal distribution with this model as the mean and a standard deviation set by the jitter parameter, and this is fit to the observed \Spitzer\ flux. This can be seen in Fig. \ref{fig:spitzer}.

%%%%%%%%%%%%%%%%%%%%%%%%%%%%%%%%%%%%%%%%%%%%%%%%%%%%%%%%%%%%%%%%%%%%%%%

\begin{figure*}
    \begin{subfigure}{\textwidth}
        \centering
        \includegraphics[width=0.86\textwidth]{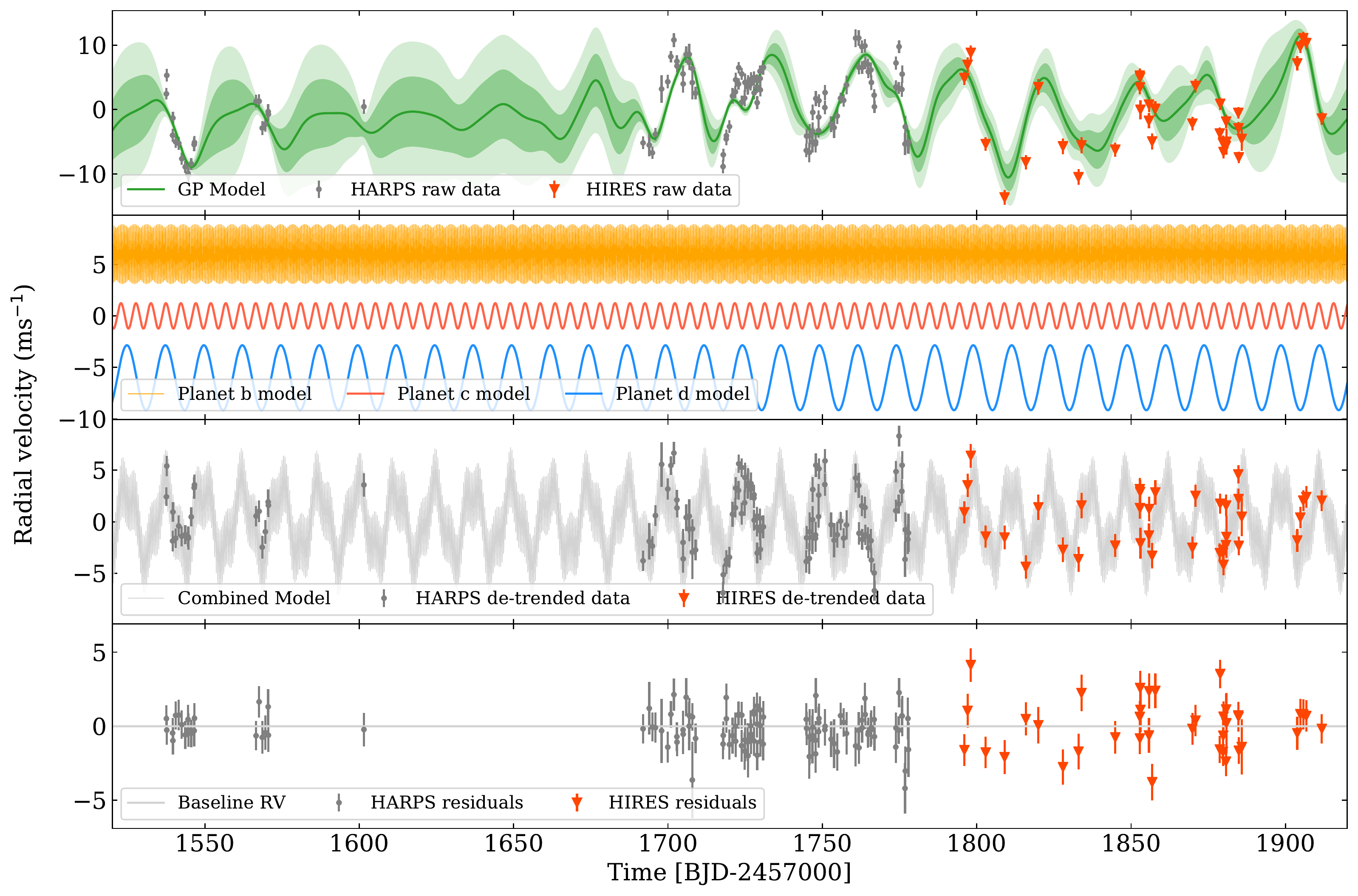}
    \end{subfigure}
    \newline
    \begin{subfigure}{\textwidth}
        \centering
        \includegraphics[width=0.86\textwidth]{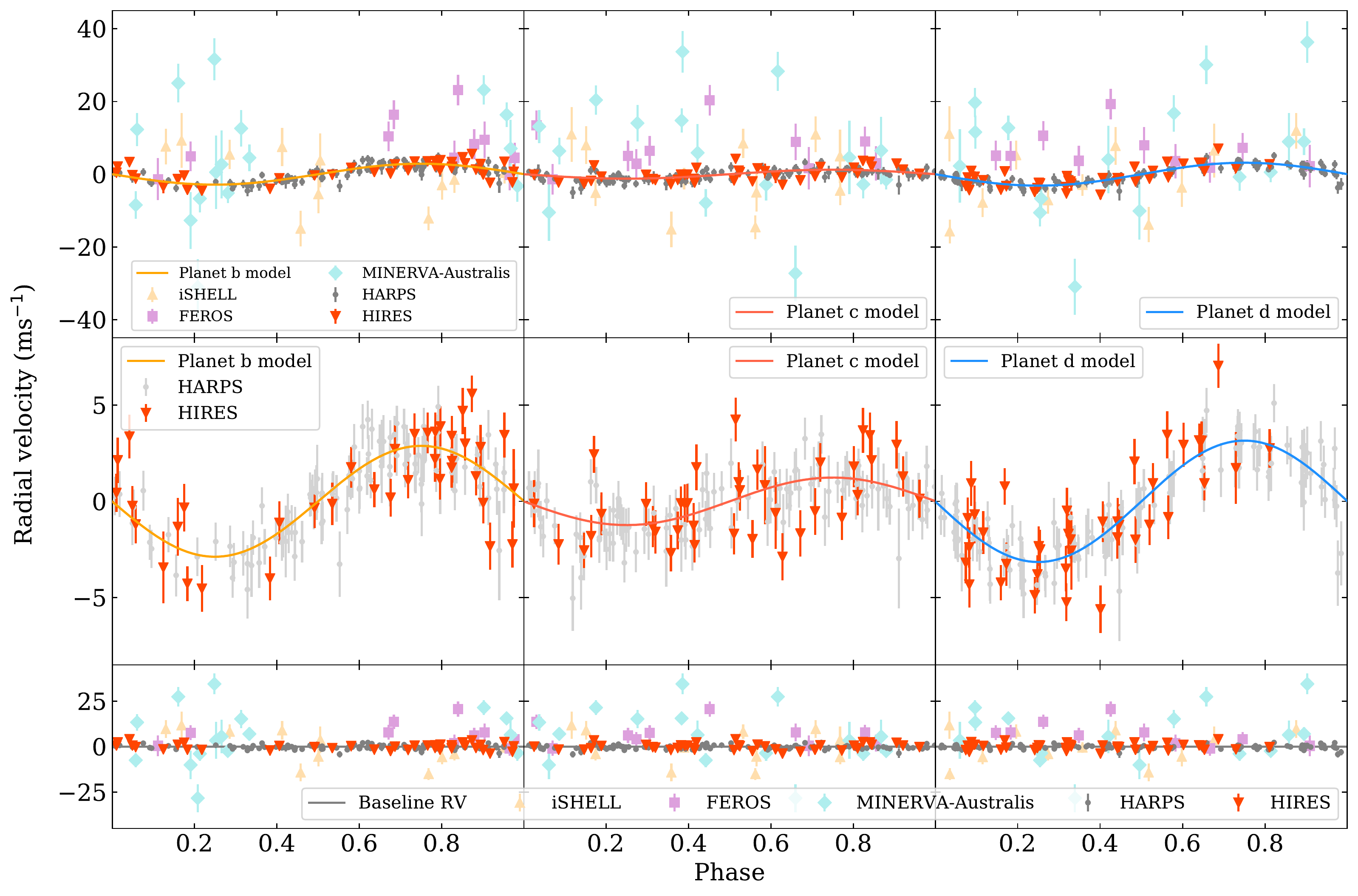}
    \end{subfigure}
    \caption{RV data plots, where the HARPS data is denoted as grey circles, HIRES as red upside down triangles, iSHELL as pale orange triangles, FEROS as pale pink squares, and M\textsc{inerva}-Australis as pale turquoise diamonds. \textbf{Top plot:} the RV data, showing the GP and planet models that have been fit. \textit{Top:} the best-fit GP used to detrend the stellar activity in the HARPS data is shown as the green line. The green shaded areas represent the 1 and 2 standard deviations of the GP fit. \textit{Upper middle:} the separate models for each planet, b (orange, offset by +6\,\ms), c (red), and d (blue, offset by -6\,\ms). \textit{Lower middle:} the total model, representing the addition of the models for planets b, c, and d, is plotted in black, and over plotted is the HARPS and HIRES data. \textit{Bottom:} the residuals after the total model, GP and offsets have been subtracted from the RV data. \textbf{Bottom plot:} the phase folds for each planet model, b (left), c (middle), and d (right), with the RV data over plotted. The top row shows all of the RV data (where the GP has been subtracted from each data set), the middle just the HARPS and HIRES data, and the bottom the residuals when the planet models have been subtracted from the RVs.}
    \label{fig:RVs}
\end{figure*}

\subsubsection{RVs} \label{sec:HARPSGP}

We do not include the iSHELL, FEROS, or M\textsc{inerva}-Australis RVs in our joint fit, as they were not found to improve the fit due to large error bars in comparison to the HARPS and HIRES data; however, they are shown to be consistent with the result of our fit (see Fig. \ref{fig:RVs}). We also do not include the archival HARPS data due to a large scatter in cadence and quality in comparison to the purpose-collected HARPS data. 

\textbf{HARPS and HIRES fitting}

In this joint fit model, we fit the HARPS and HIRES data using the same method and so they are described here in tandem. We first find predicted values of radial velocity for each planet at each \HARPS\ and HIRES timestamp using \textsc{exoplanet}. We set a wide uniform prior on $K$ for each planet, the uniform distributions centred upon $K$ values found when fitting the RV data with simple Keplerian models for all of the planets in DACE. We fit separate ``offset'' terms for HARPS and HIRES to model the systematic radial velocity, giving this a normal prior with a mean value predicted in DACE. We also fit separate ``jitter'' terms, setting wide normal priors on these, the means of which are set to double the log of the minimum error on the \HARPS\ and HIRES data respectively.

The RV data also shows significant stellar variability due to stellar rotation, and so we model this variability using another GP (see Fig. \ref{fig:RVs}, top panel of top plot). This activity can be modelled as a Quasi-Periodic signal as starspots moving across the surface of the star evolve in time and are modulated by stellar rotation. In this case, we create our own Quasi-Periodic kernel using \textsc{PyMC3}, as no such kernel is available in \textsc{exoplanet}. \textsc{PyMC3} provides a range of simple kernels\footnote{\url{https://docs.pymc.io/api/gp/cov.html}} which are easy to combine. We use their \verb|Periodic|:

\begin{equation}
    k(x,x') = \eta^2 \exp \left(-\frac{\sin^2(\pi |x-x'| \frac{1}{T})}{2l_p^2} \right),
\end{equation}

\noindent and \verb|ExpQuad| (squared exponential):

\begin{equation}
    k(x,x') = \eta^2 \exp \left(-\frac{(x-x')^2}{2l_e^2} \right)
\end{equation}

\noindent kernels. The hyperparameters are $\eta$ (the amplitude of the GP), $T$ (the recurrence timescale, equivalent to the $P_{rot}$ of the star), $l_p$ (the smoothing parameter), and $l_e$ (the timescale for growth and decay of active regions) \citep[see e.g.][]{Rasmussen2006,Haywood2014,Grunblatt2015}. 

We multiply these kernels together to create our final Quasi-Periodic kernel:

\begin{equation}
    k(x,x') = \eta^2 \exp \left(-\frac{\sin^2(\pi |x-x'| \frac{1}{T})}{2l_p^2} - \frac{(x-x')^2}{2l_e^2} \right).
\end{equation}

We use the same GP to fit the \HARPS\ and HIRES data together using the same hyperparameters. We use a normal distribution with a mean equal to the rotation period of the star found by WASP-South (see Section \ref{sec:stellaractivity} and Table \ref{tab:stellarparams}) to set a wide prior on $T$.

To bring everything together, we add the predicted radial velocities together with the offsets, and subtract these from their respective observed radial velocity values. This is then used as the prior on the GP, which is also given a noise term that is equal to an addition of the jitters with the squared error on the RV data. 

\subsection{Fit results} \label{sec:fitresults}

We first use \verb|exoplanet| to maximise the log probability of the \verb|PyMC3| model. We then use the fit parameter values this obtains as the starting point of the \verb|PyMC3| sampler, which draws samples from the posterior using a variant of Hamiltonian Monte Carlo, the No-U-Turn Sampler (NUTS). By examining the chains from earlier test runs of the model, we allow for 1000 burn-in samples which are discarded, and 5000 steps with 15 chains. 

We present our best fit parameters for the TOI-431 system from our joint fit in Table \ref{tab:planetparams}. TOI-431\,b is a super-Earth with a mass of $3.07^{+0.35}_{-0.34}$\,\mearth and a radius of 1.28 $\pm$ 0.04\,\rearth, and from this we can infer a bulk density of $7.96^{+1.05}_{-0.99}$\,\gc. This puts TOI-431\,b below the radius gap, and it is likely a stripped core with no gaseous envelope. A period of 0.49\,days puts TOI-431\,b in the rare Ultra-Short Period (USP) planet category (defined simply as planets with $P < 1$\,day); examples of systems which have USP planets include Kepler-78 \citep{Winn2018}, WASP-47 \citep{Becker2015}, and 55 Cancri \citep{Dawson2010}.
TOI-431\,c has a minimum mass of $2.83^{+0.41}_{-0.34}$\,\mearth, but the lack of transits does not allow us to fit a radius. We can use the mass-radius relation via \textsc{forecaster} \citep{Chen2017} to estimate a radius of $1.44^{+0.60}_{-0.34}$\,\rearth, which would place this planet as another super-Earth.
TOI-431\,d is a sub-Neptune with a mass of $9.90^{+1.53}_{-1.49}$\,\mearth and a radius of $3.29^{+0.09}_{-0.08}$\,\rearth, implying a bulk density of $1.36^{0.25}_{-0.24}$\,\gc. This lower density implies that TOI-431\,d probably has a gaseous envelope. We further analyse these planets in the following section.

\begin{center}
\begin{table*}
    \caption{The parameters for the planets TOI-431\,b, c, and d, calculated from our joint fit model described fully in Section \ref{sec:jointfit}. The values are given as the median values of our samples, and the uncertainties are given as the 16th and 84th percentiles. The bulk densities are then calculated using the masses and radii, assuming a spherical planet of uniform density. A calculation of the radius of TOI-431\,c can be found in Section \ref{sec:fitresults}, and discussion of the inclinations of the planets can be found in Section \ref{sec:disc}. The equilibrium temperature is calculated assuming an albedo of zero. Further joint fit model parameters to those presented here can be found in Appendix A.}
    \label{tab:planetparams}
    \begin{tabularx}{\textwidth}{ l >{\raggedright\arraybackslash}X >{\raggedright\arraybackslash}X >{\raggedright\arraybackslash}X }
    \hline
    \textbf{Parameter}                              & \textbf{TOI-431\,b}           & \textbf{TOI-431\,c}       & \textbf{TOI-431\,d}       \\
    \hline
    Period $P$ (days)                               & $0.490047^{+0.000010}_{-0.000007}$    & $4.8494^{0.0003}_{-0.0002}$               & 12.46103 $\pm$ 0.00002 \\
    Semi-major axis $a$ (AU)                        & $0.0113^{+0.0002}_{-0.0003}$          & 0.052 $\pm$ 0.001                         & 0.098 $\pm$ 0.002     \\
    Ephemeris $t_0$ (BJD-2457000)                   & $1627.538^{+0.003}_{-0.002}$          & $1625.9 \pm 0.1$                          & $1627.5453 \pm 0.0003$ \\
    Radius $R_p$ (\rearth)                          & 1.28 $\pm$ 0.04                       & -                                         & $3.29 \pm 0.09$           \\
    Impact parameter $b$                            & $0.34^{+0.07}_{-0.06}$                & -                                         & $0.15^{+0.12}_{-0.10}$    \\
    Inclination $i$ (degrees)                       & $84.3^{+1.1}_{-1.3}$                  & $< 86.35^{+0.04}_{-0.09}$                 & $89.7 \pm 0.2$  \\
    Eccentricity $e$                                & 0 (fixed)                             & 0 (fixed)                                 & 0 (fixed)         \\
    The argument of periastron $\omega$             & 0 (fixed)                             & 0 (fixed)                                 & 0 (fixed)              \\
    Radial velocity semi-amplitude $K$ ($ms^{-1}$)  & 2.88 $\pm$ 0.30                       & $1.23^{+0.17}_{-0.14}$                    & $3.16 \pm 0.46$        \\
    Mass $M_p$ (\mearth)                            & $3.07 \pm 0.35$                       & $2.83^{+0.41}_{-0.34}$ ($M \sin i$)       & $9.90^{+1.53}_{-1.49}$ \\
    Bulk density $\rho$ (\gccc)                     & $8.0 \pm 1.0$                         & -                                         & $1.36 \pm 0.25$  \\
    Equilibrium temperature $T_{eq}$ (K)            & $1862 \pm 42$                         & $867 \pm 20$                              & 633 $\pm$ 14              \\
    \hline
    \end{tabularx}
\end{table*}
\end{center}

% \section{Results} \label{sec:results}
% \input 3.Results/results.tex

\section{Discussion} \label{sec:disc}
%\subsection{Atmospheric evolution of TOI-431 b and d} % George King

\begin{figure}
    \centering
    \includegraphics[width=\columnwidth]{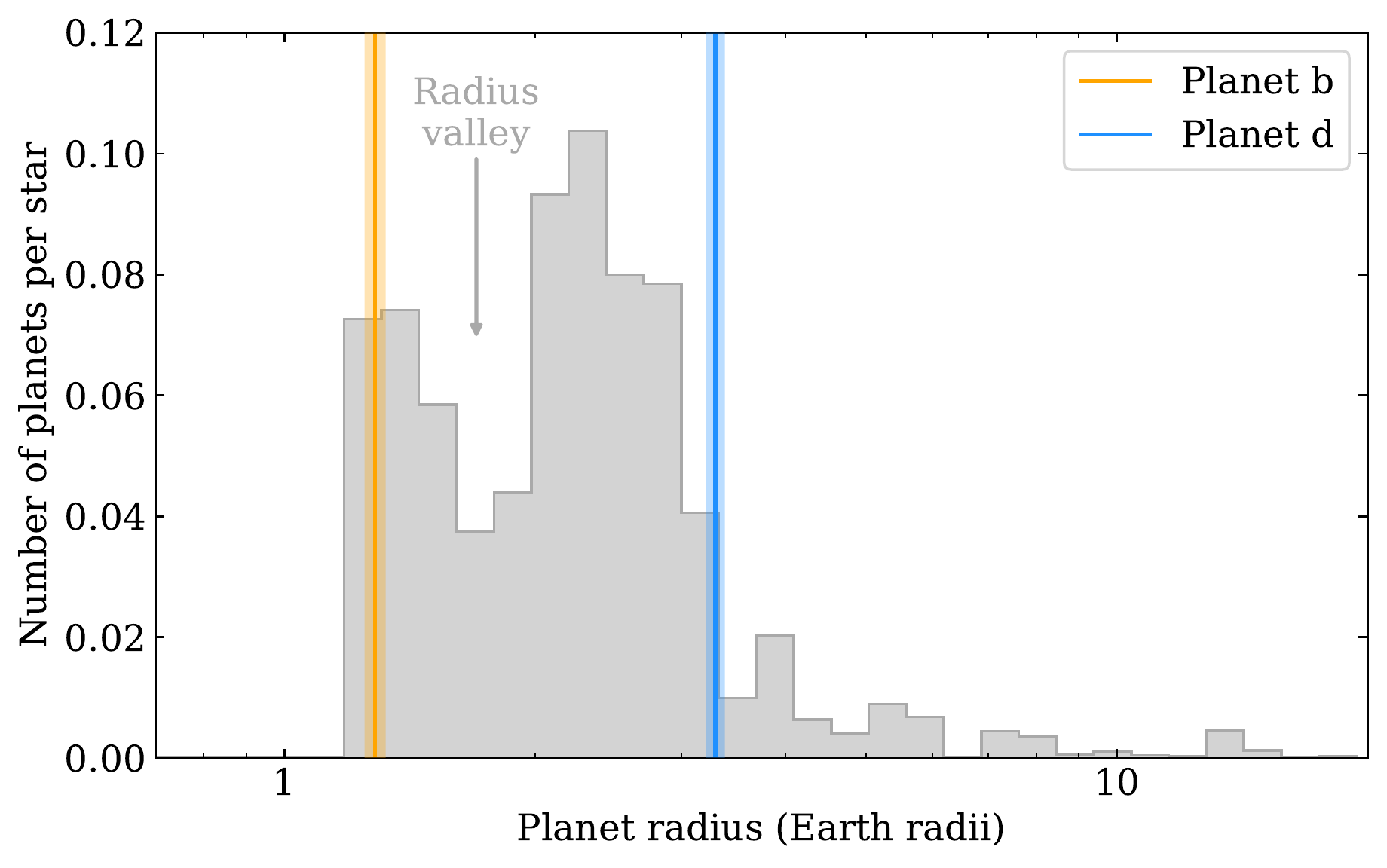}
    \caption{A histogram of planet radius for planets with orbital periods less than 100 days, as given in \citet{Fulton2018}. The radius valley can be seen at 1.7\,\rearth: below the gap are rocky super-Earths, above the gap are gaseous sub-Neptunes. TOI-431\,b (orange, with 1\,$\sigma$ confidence intervals shaded) is the former, while TOI-431\,d (blue) is the latter.}
    \label{fig:radiusvalley}
\end{figure}

The architecture of this system is unusual in that the middle planet, TOI-431\,c, is non-transiting, while the inner and outer planets are both seen to transit. Examples of this can be seen in Kepler-20 \citep{Buchhave2016}, a 6-planet system where the fifth planet out from the star does not transit, but the sixth does, and HD\,3167 \citep{Vanderburg2016,Gandolfi2017,Christiansen2017}, a 3-planet system where the middle planet does not transit as is the case with TOI-431. Using the impact parameter $b$ from Table \ref{tab:planetparams}, we calculate inclinations for TOI-431\,b and d of $(84.5^{+1.1}_{-1.3})^{\circ}$ and $89.7 \pm 0.2^{\circ}$, respectively (Table \ref{tab:planetparams}). We can calculate a limit on the inclination for TOI-431\,c assuming $b = 1$, which results in an inclination that must be $< (86.35^{+0.04}_{-0.09})^{\circ}$ in order for TOI-431\,c to be non-transiting.

The TOI-431 system is a good target system for studying planetary evolution. TOI-431\,b and d reside either side of the radius-period valley described in \citet{Fulton2017,Fulton2018,VanEylen2018} (see Fig. \ref{fig:radiusvalley}), providing a useful test-bed for the theorised mechanisms behind it. X-ray and EUV-driven photoevaporation is one of the two main proposed mechanisms \citep{Owen2017}, and we investigated its effect both now and in the past in the TOI-431 system. As no direct X-ray observations of the system exist, we had to make use of empirical formulae for relating the ratio of the X-ray and bolometric luminosities to age \citep{Jackson2012} and Rossby number (related to $P_{\rm rot}$, \citet{Wright2011,Wright2018}). We extrapolate to the EUV using the relations of \citet{King2018}. Under the assumption of energy-limited escape \citep{Watson1981,Erkaev2007}, we estimate a current mass loss rate for TOI-431\,d between $5\times 10^8$ and $5\times 10^9$ g\,s$^{-1}$. The same assumptions yield a current rate of $10^{10}$ to $10^{11}$ g\,s$^{-1}$ for TOI-431\,b, but since that planet is unlikely to retain much, if any, atmosphere, the likely true rate is much lower.

Integrating the \citet{Jackson2012} relations across the lifetime of the star, and again assuming energy-limited escape, lifetime-to-date mass loss estimates of 44 per cent and 1.0 per cent for TOI-431\,b and d respectively are found. Adding 2 per cent  extra mass and doubling the radius to account for a primordial envelope around TOI-431\,b raises the lifetime loss to 94 per cent. Again, the true value will be lower as XUV photoevaporation will not affect the rocky core, but rather the estimates calculated here demonstrate TOI-431\,b would easily have lost a typical envelope with a mass fraction of a few per cent. The value for TOI-431\,d is consistent with the density of the planet, which suggests it retains a substantial envelope.

%\subsection{Interior characterisation of TOI-431 b and d} % Jon Otegi

In order to characterize the composition of TOI-431\,b and TOI-431\,d, we model the interior considering a pure-iron core, a silicate mantle, a pure-water layer, and a H-He atmosphere. The models follow the basic structure model of \citet{Dorn2017}, with the equation of state (EOS) of the iron core taken from \citet{Hakim2018}, the EOS of the silicate-mantle from \citet{Connolly2009}, and SCVH \citep{Saumon1995} for the H-He envelope assuming protosolar composition. For water we use the QEOS of \citet{Vazan2013} for low pressures and the one of \citet{Seager2007} for pressures above 44.3\,GPa.  

\begin{figure}
\centering
\includegraphics[width=\columnwidth]{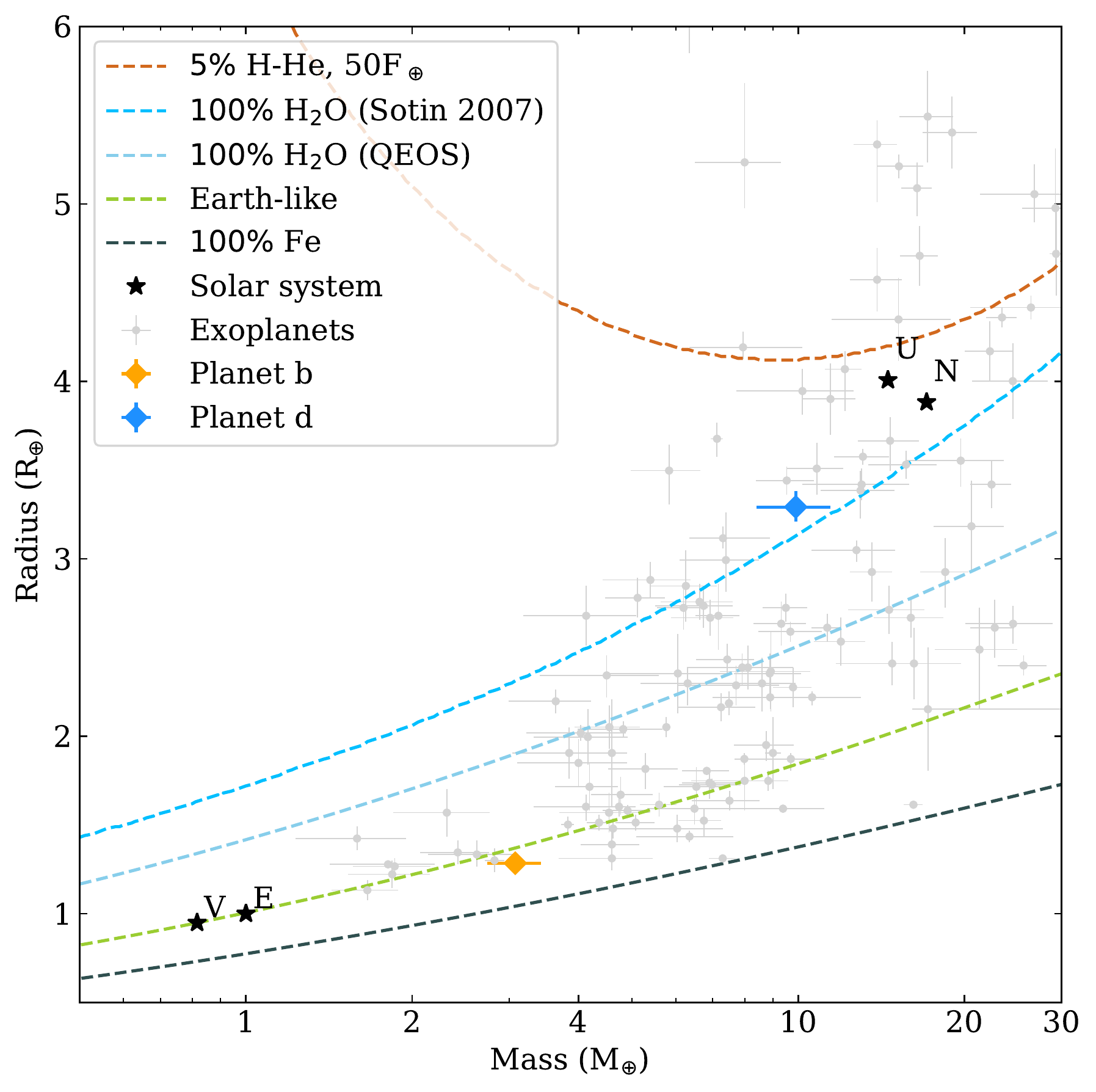}
\caption{Mass-radius diagram of known exoplanets with mass determinations better than 4$\sigma$ from the NASA exoplanet archive (\url{https://exoplanetarchive.ipac.caltech.edu}, as of 22 September 2020) shown in grey. TOI-431\,b (orange) and d (blue) are denoted as diamonds, and the Solar System planets Venus (V), Earth (E), Uranus (U), and Neptune (N) are marked as black stars. Also shown are the composition lines of iron (dark grey), Earth-like (green), and pure-water planets (pale blue and mid blue, using QEOS and EOS from \citet{Sotin2007} respectively), plus an additional line representing a planet with a 95 per cent water and a 5 per cent H-He envelope with $F/F_{\oplus}$= 50, comparable to the case of the TOI-431 planets (brown).}
\label{fig:MRplot}
\end{figure}

Fig. \ref{fig:MRplot} shows M-R curves tracing compositions of pure-iron, Earth-like, pure-water and a planet with 95 per cent water and 5 per cent H-He atmosphere subjected to a stellar radiation of $F/F_{\oplus}$= 50 (comparable to the case of the TOI-431 planets), and exoplanets with accurate and reliable mass and radius determinations. It should be noted that the position of the water line in the diagram is very sensitive to used EOS \cite[e.g.][]{Haldemann2020}. Fig. \ref{fig:MRplot} shows two water lines using QEOS and EOS from \cite{Sotin2007}. As shown in Fig. \ref{fig:MRplot}, TOI-431\,b is one of the many super-Earths following the Earth-like composition line. This suggests that it is mostly made of refractory materials. TOI-431\,d, instead, sits above the two the pure-water curves and below the 5 per cent curve, implying that the H-He mass fraction is unlikely to exceed a few per cent. Its density is lower than most of the observed sub-Neptunes. There are three planets in the catalogue presented in \citet{Otegi2020} with masses below 10\,M$_{\oplus}$ and radii above 3\,R$_{\oplus}$ (Kepler-11\,d,e and Kepler-36\,c), and all of their masses have been determined with TTVs. As shown in \citet{Otegi2020b}, reducing the uncertainties in this M-R regime would lead to significant improvements on the determination of the volatile envelope mass. As TOI-431 is in the ESPRESSO GTO target list, more observations will help to further constrain the internal structure of TOI-431\,d.

We then quantify the degeneracy between the different interior parameters and produce posterior probability distributions using a generalised Bayesian inference analysis with a Nested Sampling scheme \citep[e.g.][]{Buchner2014}. The interior parameters that are inferred include the masses of the pure-iron core, silicate mantle, water layer and H-He atmospheres. For the analysis, we use the stellar Fe/Si and Mg/Si ratios as a proxy for the planet. Table \ref{tab:interior} lists the inferred mass fractions of the core, mantle, water-layer and H-He atmosphere from the interior models. 
It should be noted, however, that our estimates have rather  large uncertainties. Indeed, in this regime of the M-R relation there is a large degeneracy, and therefore the mass ratio between the planetary layers is not well-constrained. Nevertheless, we find that TOI-431\,b has a negligible H-He envelope of 1.2x$10^{-9}\,M_{\oplus}$.

The larger companion TOI-431\,d is expected to have a significant volatile layer of H-He and/or water of about 3.6 or 33 per cent of its total mass, respectively. The nature of the volatile layer is degenerate. \\

\begin{table}
\centering
\caption{Inferred interior structure properties of TOI-431\,b and d.}
\label{tab:interior}
\begin{tabularx}{\columnwidth}{l l l}
\hline
\textbf{Interior Structure:} & \textbf{TOI-431\,b}  & \textbf{TOI-431\,d}       \\
\hline
$M_{\rm{core}}/M_{\rm{total}}$    & $0.51^{+0.15}_{-0.14}$    & $0.29^{+0.16}_{-0.13}$   \\
$M_{\rm{mantle}}/M_{\rm{total}}$  & $0.37^{+0.27}_{-0.18}$    & $0.34^{+0.23}_{-0.12}$    \\
$M_{\rm{water}}/M_{\rm{total}}$   & $0.15^{+0.12}_{-0.09}$    & $0.33^{+0.21}_{-0.15}$    \\
$M_{\rm{H-He}}/M_{\rm{total}}$    & -                         & $0.036^{+0.012}_{-0.009}$ \\
\hline
\end{tabularx}
\end{table}

%% follow-up considerations

Considering the future observation prospects of this system, for TOI-431\,d we calculate a transmission spectroscopy metric \citep[TSM;][]{Kempton2018} of $215\pm58$, after propagating the uncertainties on all system parameters. The relatively large uncertainty is dominated by the uncertainty on the planet's mass; nonetheless, this TSM value indicates that TOI-431\,d is likely among the best transmission spectroscopy targets known among small, cool exoplanets \citep[$<4 R_\oplus$, $<1000$~K; see Table 11 of][]{Guo2020}.

\section{Conclusion}
We have presented here the discovery of three new planets from the \TESS\ mission in the TOI-431 system. Our analysis is based upon 2-min cadence \TESS\ observations from 2 sectors, ground-based follow-up from LCOGT and NGTS, and space-based follow-up from \Spitzer. The photometric data was modelled jointly with RV data from the HARPS spectrograph, and further RVs from iSHELL, FEROS, and M\textsc{inerva}-Australis are included in our analysis. We find evidence to suggest that the host star is rotating with a period of 30.5 days, and account for this in our joint-fit model. Nearby contaminating stellar companions are ruled out by multiple sources of high resolution imaging. 

TOI-431\,b is a super-Earth characterised by both photometry and RVs, with an ultra-short period of 0.49 days. It likely has a negligible envelope due to substantial atmosphere evolution via photoevaporation, and an Earth-like composition.

TOI-431\,c is found in the HARPS RV data and is not seen to transit. It has a period of 4.84 days and a minimum mass similar to the mass of TOI-431\,b; extrapolating this minimum mass to a radius via the MR relation places it as a likely second super-Earth.

TOI-431\,d is a sub-Neptune with a period of 12.46 days, characterised by both photometry and RVs. It has likely retained a substantial H-He envelope of about 4 per cent of its total mass. Additionally, TOI-431\,b and d contribute to the \TESS\ Level-1 mission requirement. 

This system is a candidate for further study of planetary evolution, with TOI-431\,b and d either side of the radius valley. The system is bright, making it amenable to follow-up observations. TOI-431\,b, in particular, would potentially be an interesting target for phase-curve observations with JWST.

\section*{Acknowledgements} 
%TESS
This paper includes data collected by the \TESS\ mission. Funding for the \TESS\ mission is provided by the NASA Explorer Program. Resources supporting this work were provided by the NASA High-End Computing (HEC) Program through the NASA Advanced Supercomputing (NAS) Division at Ames Research Center for the production of the SPOC data products.

%TESS SPOC
We acknowledge the use of public \TESS\ Alert data from pipelines at the \TESS\ Science Office and at the \TESS\ Science Processing Operations Center.

%ESO
This study is based on observations collected at the European Southern Observatory under ESO programme 1102.C-0249 (PI: Armstrong). Additionally the archival HARPS data dating from 2004 to 2015 were obtained under the following programmes: 072.C-0488 (PI: Mayor); 183.C-0972 (PI: Udry); 085.C-0019 (PI: Lo Curto); 087.C-0831 (PI: Lo Curto); and 094.C-0428 (PI: Brahm).

%iSHELL
This work was supported by grants to P.P. from NASA (award 18-2XRP18\_2-0113), the National Science Foundation (Astronomy and Astrophysics grant 1716202), the Mount Cuba Astronomical Foundation, and George Mason University start-up funds.The NASA Infrared Telescope Facility is operated by the University of Hawaii under contract NNH14CK55B with NASA. 

% NGTS
This paper is in part based on data collected under the NGTS project at the ESO La Silla Paranal Observatory.  The NGTS facility is operated by the consortium institutes with support from the UK Science and Technology Facilities Council (STFC)  projects ST/M001962/1 and  ST/S002642/1.

%LCOGT
This work makes use of observations from the LCOGT network.

% Minerva
M\textsc{inerva}-Australis is supported by Australian Research Council LIEF Grant LE160100001, Discovery Grant DP180100972, Mount Cuba Astronomical Foundation, and institutional partners University of Southern Queensland, UNSW Sydney, MIT, Nanjing University, George Mason University, University of Louisville, University of California Riverside, University of Florida, and The University of Texas at Austin.

We respectfully acknowledge the traditional custodians of all lands throughout Australia, and recognise their continued cultural and spiritual connection to the land, waterways, cosmos, and community. We pay our deepest respects to all Elders, ancestors and descendants of the Giabal, Jarowair, and Kambuwal nations, upon whose lands the M\textsc{inerva}-Australis facility at Mt Kent is situated.

%Spitzer
This work is based in part on observations made with the {\it Spitzer} Space Telescope, which was operated by the Jet Propulsion Laboratory, California Institute of Technology under a contract with NASA. Support for this work was provided by NASA through an award issued by JPL/Caltech.

% Gemini 
Based on observations obtained at the international Gemini Observatory, a program of NSF’s NOIRLab acquired through the Gemini Observatory Archive at NSF’s NOIRLab, which is managed by the Association of Universities for Research in Astronomy (AURA) under a cooperative agreement with the National Science Foundation on behalf of the Gemini Observatory partnership: the National Science Foundation (United States), National Research Council (Canada), Agencia Nacional de Investigación y Desarrollo (Chile), Ministerio de Ciencia, Tecnología e Innovación (Argentina), Ministério da Ciência, Tecnologia, Inovações e Comunicações (Brazil), and Korea Astronomy and Space Science Institute (Republic of Korea). Data collected under program GN-2019A-LP-101.  

% Mauna Kea (for Gemini)
This work was enabled by observations made from the Gemini North telescope, located within the Maunakea Science Reserve and adjacent to the summit of Maunakea. We are grateful for the privilege of observing the Universe from a place that is unique in both its astronomical quality and its cultural significance.

%SOAR
This work is based in part on observations obtained at the Southern Astrophysical Research (SOAR) telescope, which is a joint project of the Minist\'{e}rio da Ci\^{e}ncia, Tecnologia e Inova\c{c}\~{o}es (MCTI/LNA) do Brasil, the US National Science Foundation’s NOIRLab, the University of North Carolina at Chapel Hill (UNC), and Michigan State University (MSU).

%%% DATABASES, SOFTWARE ETC
% exoplanet
This research made use of \textsf{exoplanet} \citep{exoplanet:exoplanet} and its dependencies \citep{exoplanet:agol19, exoplanet:astropy13, exoplanet:astropy18, exoplanet:exoplanet, exoplanet:kipping13, exoplanet:luger18, exoplanet:pymc3, exoplanet:theano}.

% DACE
This publication makes use of The Data \& Analysis Center for Exoplanets (DACE), which is a facility based at the University of Geneva (CH) dedicated to extrasolar planets data visualisation, exchange and analysis. DACE is a platform of the Swiss National Centre of Competence in Research (NCCR) PlanetS, federating the Swiss expertise in Exoplanet research. The DACE platform is available at https://dace.unige.ch.

% GAIA
This work has made use of data from the European Space Agency (ESA) mission {\it Gaia} (\url{https://www.cosmos.esa.int/gaia}), processed by the {\it Gaia} Data Processing and Analysis Consortium (DPAC, \url{https://www.cosmos.esa.int/web/gaia/dpac/consortium}). Funding for the DPAC has been provided by national institutions, in particular the institutions participating in the {\it Gaia} Multilateral Agreement.

%%% Authors
AO is supported by an STFC studentship. 
DJA acknowledges support from the STFC via an Ernest Rutherford Fellowship (ST/R00384X/1).
RB acknowledges support from FONDECYT Post-doctoral Fellowship Project 3180246, and from the Millennium Institute of Astrophysics (MAS).
IJMC acknowledges support from the NSF through grant AST-1824644.
AJ acknowledges support from FONDECYT project 1210718, and from ANID – Millennium Science Initiative – ICN12\_009.
MNG. acknowledges support from MIT's Kavli Institute as a Juan Carlos Torres Fellow.
JSJ acknowledges support by FONDECYT grant 1201371 and partial support from CONICYT project Basal AFB-170002.
We acknowledge support by FCT - Funda\c{c}\~ao para a Ci\^encia e a Tecnologia through national funds and by FEDER through COMPETE2020 - Programa Operacional Competitividade e Internacionaliza\c{c}\~ao by these grants: UID/FIS/04434/2019; UIDB/04434/2020; UIDP/04434/2020; PTDC/FIS-AST/32113/2017 \& POCI-01-0145-FEDER-032113; PTDC/FIS-AST/28953/2017 \& POCI-01-0145-FEDER-028953. 
VA, EDM and SCCB acknowledge the support from FCT through Investigador FCT contracts  IF/00650/2015/CP1273/CT0001, IF/00849/2015/CP1273/CT0003, and IF/01312/2014/CP1215/CT0004 respectively.
ODSD is supported in the form of work contract (DL 57/2016/CP1364/CT0004) funded by FCT.
CD acknowledges the SNSF Ambizione Grant 174028. SH acknowledge support by the fellowships PD/BD/128119/2016 funded by FCT (Portugal).
JKT acknowledges that support for this work was provided by NASA through Hubble Fellowship grant HST-HF2-51399.001 awarded by the Space Telescope Science Institute, which is operated by the Association of Universities for Research in Astronomy, Inc., for NASA, under contract NAS5-26555.
DD acknowledges support through the TESS Guest Investigator Program Grant 80NSSC19K1727.
JLB and DB have been supported by the Spanish State Research Agency (AEI) Project  No. MDM-2017-0737 Unidad de Excelencia “María de Maeztu”- Centro de Astrobiología (CSIC/INTA).
SH acknowledges CNES funding through the grant 837319.
PJW acknowledges support from STFC through consolidated grants ST/P000495/1 and ST/T000406/1.
L.M.W. is supported by the Beatrice Watson Parrent Fellowship and NASA ADAP Grant 80NSSC19K0597.

\section*{Data Availability}
The TESS data are available from the Mikulski Archive for Space Telescopes (MAST), at \url{https://heasarc.gsfc.nasa.gov/docs/tess/data-access.html}. The other photometry from the LCOGT, NGTS, and \Spitzer\, as well as all of the RV data, are available for public download from the ExoFOP-TESS archive at \url{https://exofop.ipac.caltech.edu/tess/target.php?id=31374837}. This data is labelled ``Osborn+ 2021'' in their descriptions. The high-resolution imaging data is also available from the ExoFOP TESS archive. The model code underlying this article will be shared on reasonable request to the corresponding author.

%%%%%%%%%%%%%%%%%%%%%%%%%%%%%%%%%%%%%%%%%%%%%%%%%%



\bibliographystyle{mnras}
\bibliography{bib} 

%%%%%%%%%%%%%%%%%%%%%%%%%%%%%%%%%%%%%%%%%%%%%%%%%%

%%%%%%%%%%%%%%%%% APPENDICES %%%%%%%%%%%%%%%%%%%%%
\appendix

\section{Further Joint Fit Parameters}
Further parameters from our joint fit model (described in Section \ref{sec:jointfit}) are presented in Table \ref{tab:furtherfitparams}.
\begin{table*} 
\caption{Further parameters to those presented in Table \ref{tab:planetparams}: the prior distributions input into our joint fit model (described fully in Section \ref{sec:jointfit}), and the fit values resulting from the model. The priors are created using distributions in \textsc{PyMC3}, and the relevant inputs to each distribution are listed. The fit values are given as the median values of our samples, and the uncertainties are given as the 16th and 84th percentiles. Where necessary, the specific planet a parameter is describing is noted in square brackets.}
\label{tab:furtherfitparams}
\adjustbox{valign=t}{\begin{minipage}{0.5\textwidth}%
\begin{tabularx}{\textwidth}{X p{3cm} X}
    \toprule
    \textbf{Parameter} & \textbf{Prior Distribution} & \textbf{Fit Value} \\
    \midrule
    \textbf{Planets} & & \\
    Period $P$ [b] (days)                  & $\mathcal{N}(0.4900657, 0.001)$       & $0.490047^{+0.000010}_{-0.000007}$\\
    Period $P$ [c] (days)                  & $\mathcal{N}(4.849427, 0.1)$          & $4.8494^{+0.0003}_{-0.0002}$\\
    Period $P$ [d] (days)                  & $\mathcal{N}(12.46109, 0.01)$         & $12.46103 \pm 0.00002$\\
    Ephemeris $t_0$ [b] \newline (BJD-2457000)      & $\mathcal{N}(1627.533, 0.1)$ & $1627.538^{+0.003}_{-0.002}$\\
    Ephemeris $t_0$ [c] \newline (BJD-2457000)      & $\mathcal{N}(1625.888, 0.1)$ & $1625.87 \pm 0.10$\\
    Ephemeris $t_0$ [d] \newline (BJD-2457000)      & $\mathcal{N}(1627.545, 0.1)$ & $1627.5453 \pm 0.0003$\\
    $\log{(R_p)}$ [b] (\rsun)    & $\mathcal{N}(-4.35^*, 1.0)$    & $-4.44 \pm 0.03$\\
    $\log{(R_p)}$ [d] (\rsun)    & $\mathcal{N}(-3.41^*, 1.0)$    & $-3.50 \pm 0.03$\\
    \midrule
    \textbf{Star} & & \\
    Mass (\msun)                        & $\mathcal{N_T}(0.77, 0.7, 0.0, 3.0)$      & $0.81 \pm 0.05$  \\
    Radius (\rsun)                      & $\mathcal{N_T}(0.729, 0.022, 0.0, 3.0)$   & $0.72 \pm -0.02$ \\
    \midrule
    \textbf{\textit{TESS}} & & \\
    Mean                                & $\mathcal{N}(0.0, 1.0)$      & $0.00006 \pm 0.00006$   \\
    GP $\log{(s2)}$                     & $\mathcal{N}(-15.257^{\dagger}, 0.1)$  & $-15.539 \pm 0.008$     \\
    GP $\log{(w0)}$                     & $\mathcal{N}(0.0, 0.1)$      & $0.19 \pm 0.08$    \\
    GP $\log{(Sw4)}$                    & $\mathcal{N}(-15.257^{\dagger}, 0.1)$  & {$-15.37 \pm 0.09$}  \\
    \midrule
    \textbf{LCOGT (ingress)} & & \\
    Mean                                & $\mathcal{N}(0.0, 1.0)$  & $-0.00044 \pm 0.00008$ \\
    \midrule
    \textbf{LCOGT (egress)} & &  \\
    Mean                                & $\mathcal{N}(0.0, 1.0)$  & $0.00002 \pm 0.00006$ \\
    \midrule
    \textbf{NGTS} & & \\
    Mean                                & $\mathcal{N}(0.0, 1.0)$  & $-0.00015^{+0.00008}_{-0.00007}$ \\
\end{tabularx}
\end{minipage}\hfill}%
\adjustbox{valign=t}{\begin{minipage}{0.5\textwidth}%
\begin{tabularx}{\textwidth}{X p{3cm} X}
    \toprule
    \textbf{Parameter} & \textbf{Prior Distribution} & \textbf{Fit Value}  \\
    \midrule
    \textbf{\textit{Spitzer}} & & \\
    Jitter                              & $\mathcal{N}(337.0, 20.0)$   & $345 \pm 8$             \\
    Pixel coefficient $c_1$             & $\mathcal{N}(1236218, 10^5)$ & $1448286^{+68271}_{-69627}$  \\
    Pixel coefficient $c_2$             & $\mathcal{N}(468921, 10^5)$  & $408211^{+14963}_{-14570}$   \\
    Pixel coefficient $c_3$             & $\mathcal{N}(-917568, 10^5)$ & $-832924^{+62790}_{-62527}$  \\
    Pixel coefficient $c_4$             & $\mathcal{N}(465062, 10^5)$  & $428366^{+16837}_{-16824}$   \\
    Pixel coefficient $c_5$             & $\mathcal{N}(693929, 10^5)$  & $688664^{+10881}_{-10749}$   \\
    Pixel coefficient $c_6$             & $\mathcal{N}(554898, 10^5)$  & $542039^{+12467}_{-12391}$   \\
    Pixel coefficient $c_7$             & $\mathcal{N}(-205010, 10^5)$ & $-194425^{+61256}_{-59207}$  \\
    Pixel coefficient $c_8$             & $\mathcal{N}(564035, 10^5)$  & $522150^{+12762}_{-12784}$   \\
    Pixel coefficient $c_9$             & $\mathcal{N}(618285, 10^5)$  & $669652^{+22918}_{-22697}$   \\
    Time dependent \newline ramp coefficient $f$   & $\mathcal{N}(0.0, 170000)$   & $2017^{+9457}_{-9651}$ \\
    Time dependent \newline ramp coefficient $g$   & $\mathcal{N}(0.0, 170000)$   & $618^{+522}_{-518}$ \\
    Offset constant $h$                            & $\mathcal{N}(0.0, 10^4)$     & $-1543^{+3755}_{-3760}$ \\
    \midrule
    \multicolumn{3}{l}{\textbf{HARPS and HIRES}} \\
    HARPS Offset                    & $\mathcal{N}(48830.87, 10.0)$               & $48828 \pm 2$         \\
    $\log{(\rm{Jitter_{HARPS}})}$   & $\mathcal{N}(-0.2661^{\ddagger}, 5.0)$      & $-5.06^{+2.10}_{-3.37}$      \\
    HIRES Offset                    & $\mathcal{N}(0.01, 10.0)$                   & $-2.07 \pm 2.34$         \\
    $\log{(\rm{Jitter_{HIRES}})}$   & $\mathcal{N}(-0.2659^{\ddagger}, 5.0)$      & $-0.05^{+0.36}_{-0.43}$      \\
    GP recurrence \newline timescale  $T$ \newline (stellar rotation \newline period) (days) & $\mathcal{N}(30.5, 0.7)$ & $30.7 \pm 0.6$            \\
    GP amplitude $\eta$    & $\mathcal{HC}(5.0)$                   & $5.48^{+1.12}_{-0.83}$ \\
    GP lengthscale $l_e$  & $\mathcal{N_T}(30.0, 20.0, 25.0, -)$  & $31.5^{+6.3}_{-4.2}$ \\
    GP lengthscale $l_p$   & $\mathcal{N_T}(0.1, 10.0, 0.0, 1.0)$  & $0.47^{+0.10}_{-0.09}$ \\
\end{tabularx}
\end{minipage}}
\begin{tabularx}{\textwidth}{X}
    \bottomrule
\end{tabularx}
\parbox{\textwidth}{\vspace{1eX}
\textbf{Distribution descriptions:} \\
$\mathcal{N}(\mu, \sigma)$: a normal distribution with a mean $\mu$ and a standard deviation $\sigma$; \\
$\mathcal{N_B}(\mu, \sigma, a, b)$: a bounded normal distribution with a mean $\mu$, a standard deviation $\sigma$, an lower bound $a$, and an upper bound $b$ (bounds optional); \\
$\mathcal{N_T}(\mu, \sigma, a, b)$: a truncated normal distribution with a mean $\mu$, a standard deviation $\sigma$, a lower bound $a$, and an upper bound $b$ (bounds optional); \\
$\mathcal{HC}(\beta)$: a Half-Cauchy distribution with a single beta parameter $\beta$. \\
\textbf{Prior values:} \\
$^*$ equivalent to $0.5(\log{(D)}) + \log{(\rstar)}$ where $D$ is the transit depth and $\rstar$ is the value of the prior on the stellar radius ($\rsun$); \\
$^{\dagger}$ equivalent to the log of the variance of the \TESS\ flux; \\
$^{\ddagger}$ equivalent to 2 times the log of the minimum error on the HARPS or HIRES RV data, respectively.
}
\end{table*}

\section{Stellar Activity Indicators}

\begin{figure*} 
    \centering
    \begin{subfigure}{0.49\textwidth}
    \includegraphics[width=\columnwidth]{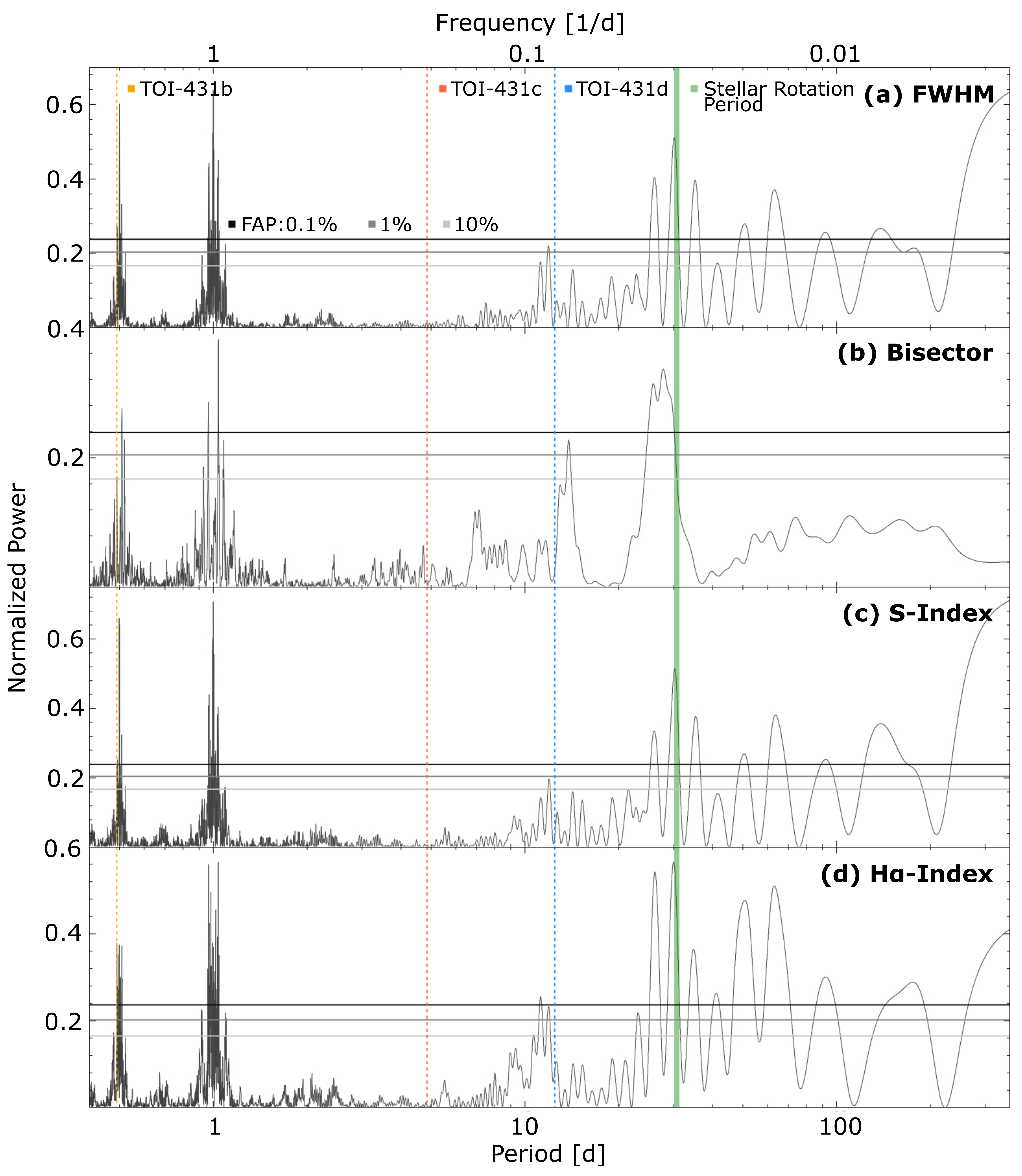}
    \end{subfigure}
    \begin{subfigure}{0.49\textwidth}
    \includegraphics[width=\columnwidth]{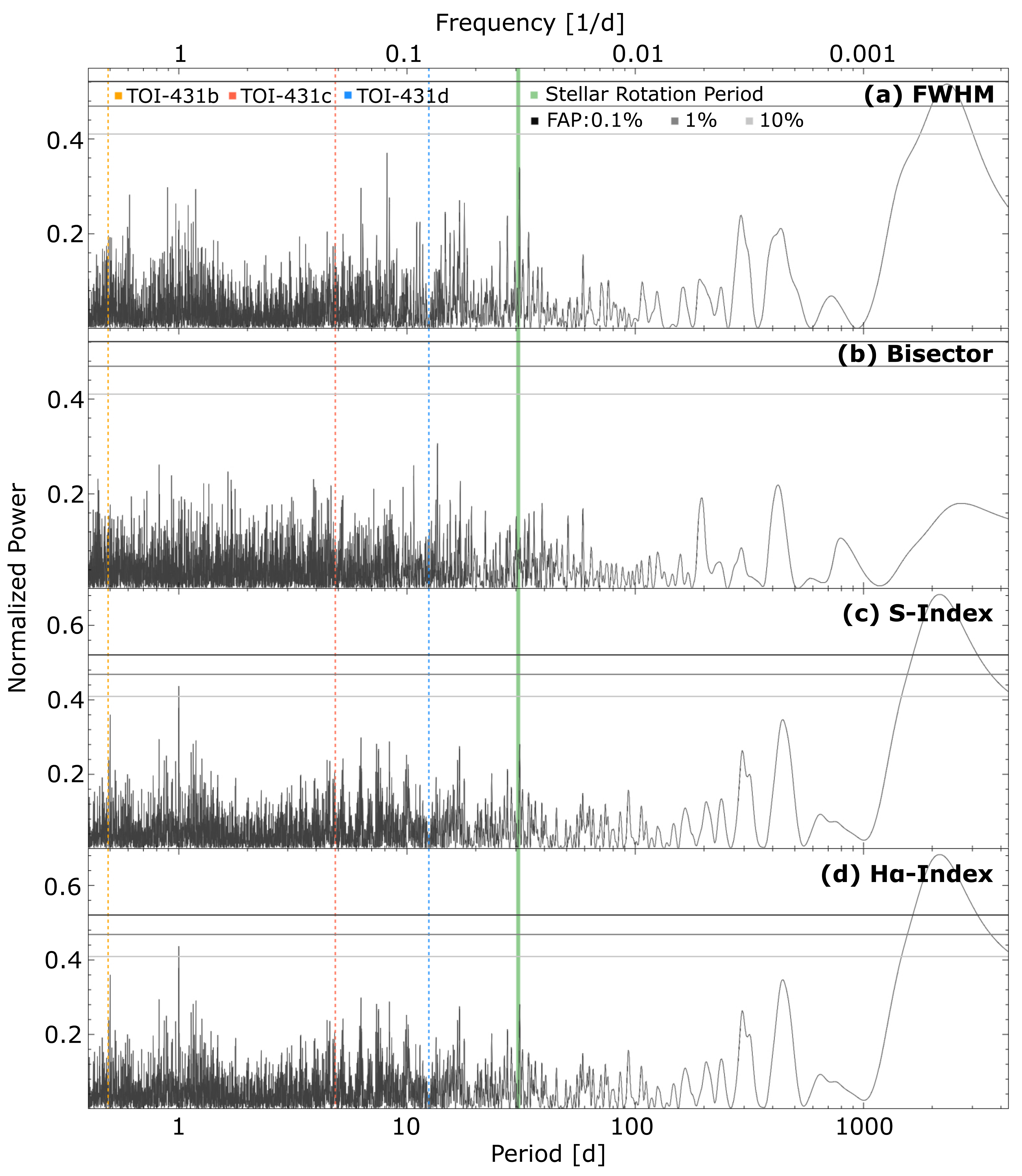}
    \end{subfigure}
    \begin{subfigure}{0.49\textwidth}
    \includegraphics[width=\columnwidth]{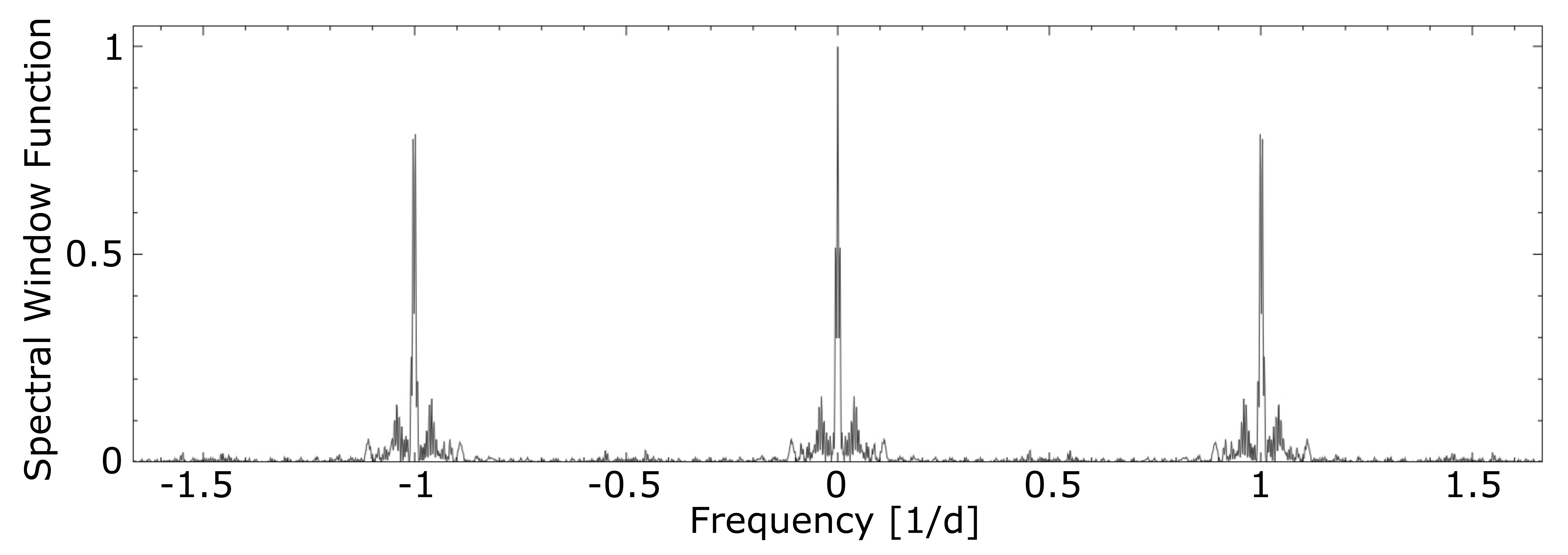}
    \end{subfigure}
    \begin{subfigure}{0.49\textwidth}
    \includegraphics[width=\columnwidth]{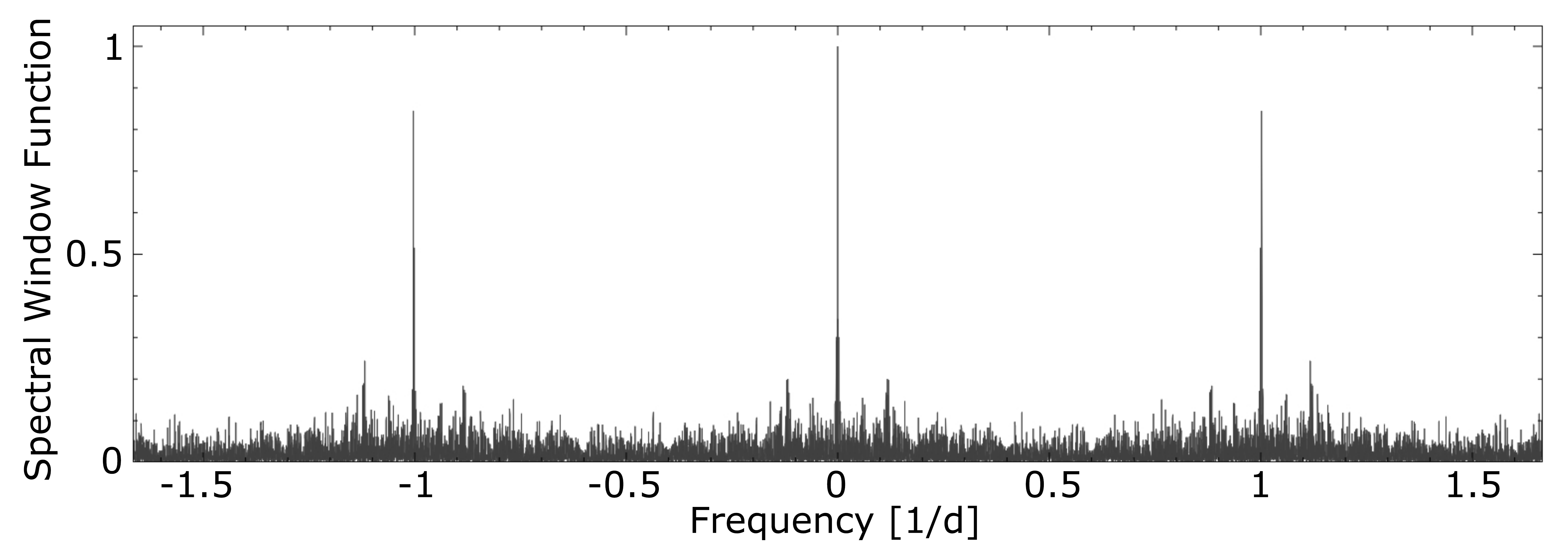}
    \end{subfigure}
    \caption{Periodograms for the activity indicators (top row) and window functions (bottom row) from the HARPS data, including the purpose-collected HARPS data from February to October 2019 (left), and the archival HARPS data from 2004 to 2015 (right), illustrating that there is no significant power at the 4.85\,day period of TOI-431\,c. The best fit periods (see Table \ref{tab:planetparams}) of TOI-431\,b (yellow), c (red), and d (blue), have been denoted by dotted lines, and the 1 standard deviation interval of the rotation period of the star has been shaded in green.}
    \label{fig:activityindicators}
\end{figure*}

Further to Fig. \ref{fig:RVlombscargle}, periodograms of stellar activity indicators for both the archival and the purpose-collected HARPS data are presented in Fig. \ref{fig:activityindicators}. 

\section{Data}

The HARPS and HIRES RV data are presented in Tables \ref{tab:dataharps} and \ref{tab:datahires}, respectively. 

\begin{table*}
	\caption{HARPS spectroscopy from February to October 2019.}
	\label{tab:dataharps}
	\begin{threeparttable}
	\begin{tabular}{lllllll}
	\hline
	Time                    & RV            & $\sigma_\textrm{RV}$  & FWHM          & Bisector      & Contrast  & S$_\textrm{MW}$ \\
	(RJD)                   & ($ms^{-1}$)   & ($ms^{-1}$)           & ($ms^{-1}$)   & ($ms^{-1}$)   &           &               \\
	\hline
	58537.53770973021       & 48830.979962  & 0.894407              & 6330.143967   & 38.148888     & 49.532876 & 0.370009      \\
	58537.655514969956      & 48833.848987  & 0.994823              & 6330.289387   & 36.923112     & 49.534854 & 0.361645      \\
	58539.53381296992       & 48824.538870  & 1.006177              & 6324.786911   & 39.072933     & 49.567807 & 0.365675      \\
	\vdots                  & \vdots        & \vdots                & \vdots        & \vdots        & \vdots    & \vdots        \\
	\end{tabular}
    \begin{tablenotes}
    \item The full HARPS data products can be found on ExoFOP-TESS at \url{https://exofop.ipac.caltech.edu/tess/target.php?id=31374837}
    \end{tablenotes}
    \end{threeparttable}
\end{table*}

\begin{table}
	\caption{HIRES spectroscopy from x to x 20xx.}
	\label{tab:datahires}
	\begin{threeparttable}
	\begin{tabular}{lll}
	\hline
	Time               & RV                 & $\sigma_\textrm{RV}$   \\
	(BJD TDB)          & ($ms^{-1}$)        & ($ms^{-1}$)            \\
	\hline
	2458796.014464     & 4.90676701782345   & 1.06348240375519     \\
	2458797.0428       & 6.94764041206104   & 1.14499938488007     \\
	2458798.095775     & 8.81269072598892   & 1.1401127576828      \\
	\vdots             & \vdots             & \vdots               \\
	\end{tabular}
    \begin{tablenotes}
    \item The full HIRES data products can be found on ExoFOP-TESS at \url{https://exofop.ipac.caltech.edu/tess/target.php?id=31374837}
    \end{tablenotes}
    \end{threeparttable}
\end{table}

\section{Author affiliations} \label{sec:affiliations}

% List of institutions
$^{1}$Department of Physics, University of Warwick, Gibbet Hill Road, Coventry CV4 7AL, UK\\
$^{2}$Centre for Exoplanets and Habitability, University of Warwick, Gibbet Hill Road, Coventry CV4 7AL, UK\\
$^{3}$George Mason University, 4400 University Drive MS 3F3, Fairfax, VA 22030, USA\\
$^{4}$Facultad de Ingeniería y Ciencias, Universidad Adolfo Ib\'a\~nez, Av.\ Diagonal las Torres 2640, Pe\~nalol\'en, Santiago, Chile\\
$^{5}$Millennium Institute for Astrophysics, Chile\\
$^{6}$Centre for Astrophysics, University of Southern Queensland, West Street, Toowoomba, QLD 4350 Australia\\
$^{7}$Division of Geological and Planetary Sciences, 1200 E California Blvd, Pasadena, CA, 91125, USA\\
$^{8}$Department of Physics and Astronomy, University of Kansas, Lawrence, KS, USA\\
$^{9}$Instituto de Astrof\'{\i}sica e Ci\^encias do Espa\c co, Universidade do Porto, CAUP, Rua das Estrelas, 4150-762, Porto, Portugal\\
$^{10}$Departamento de F\'isica e Astronomia, Faculdade de Ci\^encias, Universidade do Porto, Rua do Campo Alegre, 4169-007 Porto, Portugal\\
$^{11}$Center for Astrophysics \textbar \ Harvard \& Smithsonian, 60 Garden Street, Cambridge, MA 02138, USA \\ 
$^{12}$Banting Fellow\\
$^{13}$Leiden Observatory, Leiden University, NL-2333 CA Leiden, The Netherlands\\
$^{14}$Department of Space, Earth and Environment, Chalmers University of Technology, Onsala Space Observatory, SE-439 92 Onsala, Sweden\\
$^{15}$Astrophysics Group, Keele University, Staffordshire ST5 5BG, U.K.\\
$^{16}$NASA Ames Research Center, Moffett Field, CA 94035, USA\\
$^{17}$Depto. Astrof\'{\i}sica, Centro de Astrobiolog\'{\i}a (CSIC/INTA), ESAC Campus, 28692 Villanueva de la Ca\~nada (Madrid), SPAIN\\
$^{18}$Observatoire de l’Universit\'e de Gen\`eve, Chemin des Maillettes 51, 1290 Versoix, Switzerland\\
$^{19}$University of Z\"{u}rich, Institute for Computational Science, Winterthurerstrasse 190, 8057 Z\"{u}rich, Switzerland\\
$^{20}$Department of Physics \& Astronomy, Vanderbilt University, Nashville, TN 37235, USA\\
$^{21}$Department of Physics and Kavli Institute for Astrophysics and Space Research, Massachusetts Institute of Technology, Cambridge, MA 02139, USA\\
$^{22}$Dunlap Institute for Astronomy and Astrophysics, University of Toronto, 50 St. George Street, Toronto, Ontario M5S 3H4, Canada\\
$^{23}$Department of Earth, Atmospheric and Planetary Sciences, Massachusetts Institute of Technology, Cambridge, MA 02139, USA\\
$^{24}$Department of Aeronautics and Astronautics, MIT, 77 Massachusetts Avenue, Cambridge, MA 02139, USA\\
$^{25}$Department of Astrophysical Sciences, Peyton Hall, 4 Ivy Lane, Princeton, NJ 08540, USA\\
$^{26}$School of Physics and Astronomy, University of Leicester, LE1 7RH, UK\\
$^{27}$Department of Astronomy, University of Florida, 211 Bryant Space Science Center, Gainesville, FL, 32611, USA\\
$^{28}$University of Maryland, Baltimore County, 1000 Hilltop Circle, Baltimore, MD 21250, USA\\
$^{29}$NASA Goddard Space Flight Center, 8800 Greenbelt Road, Greenbelt, MD 20771, USA\\
$^{30}$Departement of Physics, and Institute for Research on Exoplanets, Universite de Montreal, Montreal, Canada\\
$^{31}$Department of Astronomy, The University of Texas at Austin, Austin, TX 78712, USA\\
$^{32}$Cerro Tololo Inter-American Observatory, Casilla 603, La Serena 1700000, Chile\\
$^{33}$NASA Exoplanet Science Institute, Caltech/IPAC, Mail Code 100-22, 1200 E. California Blvd., Pasadena, CA 91125, USA\\
$^{34}$International Center for Advanced Studies (ICAS) and ICIFI (CONICET), ECyT-UNSAM, Campus Miguelete, 25 de Mayo y Francia, (1650) Buenos Aires, Argentina\\
$^{35}$Department of Physics and Astronomy, University of New Mexico, 210 Yale Blvd NE, Albuquerque, NM 87106, USA\\
$^{36}$Department of Astronomy, The University of California, Berkeley, CA 94720, USA\\
$^{37}$Space Telescope Science Institute, 3700 San Martin Drive, Baltimore, MD 21218, USA\\
$^{38}$European Southern Observatory, Alonso de Cordova 3107, Vitacura, Santiago, Chile\\
$^{39}$NASA Exoplanet Science Institute/Caltech-IPAC, MC 314-6, 1200 E California Blvd, Pasadena, CA 91125, USA\\
$^{40}$Department of Earth Sciences \textbar \ University of Hawai'i at M\={a}noa, Honolulu, HI 96822, USA\\
$^{41}$Department of Physics \& Astronomy, Mississippi State University, Starkville, MS 39762, USA\\
$^{42}$Department of Astronomy \& Astrophysics, University of California, Santa Cruz, CA 95064, USA\\
$^{43}$National Science Foundation Graduate Research Fellow\\
$^{44}$Jet Propulsion Laboratory, California Institute of Technology, 4800 Oak Grove Drive, Pasadena, CA 91109, USA\\
$^{45}$Max-Planck-Institut f\"ur Astronomie, K\"onigstuhl 17, Heidelberg 69117, Germany\\
$^{46}$Department of Astronomy, California Institute of Technology, Pasadena, CA 91125, USA\\
$^{47}$Aix Marseille Univ, CNRS, CNES, LAM, Marseille, France\\
$^{48}$Institute for Astronomy, University of Hawai`i, 2680 Woodlawn Drive, Honolulu, HI 96822, USA\\
$^{49}$501 Campbell Hall, University of California at Berkeley, Berkeley, CA 94720, USA\\
$^{50}$Departamento de Astronomía, Universidad de Chile, Camino El Observatorio 1515, Las Condes, Santiago, Chile\\
$^{51}$Centro de Astrof\'isica y Tecnolog\'ias Afines (CATA), Casilla 36-D, Santiago, Chile\\
$^{52}$Dept.\ of Physics \& Astronomy, Swarthmore College, Swarthmore PA 19081, USA\\
$^{53}$Department of Earth and Planetary Sciences, University of California, Riverside, CA 92521, USA\\
$^{54}$Department of Physics and Astronomy, University of Louisville, Louisville, KY 40292, USA\\
$^{55}$Department of Physics and Astronomy, The University of North Carolina at Chapel Hill, Chapel Hill, NC 27599-3255, USA\\
$^{56}$NCCR PlanetS, Centre for Space \& Habitability, University of Bern, Bern, Switzerland\\
$^{57}$U.S. Naval Observatory, 3450 Massachusetts Avenue NW, Washington, D.C. 20392, USA\\
$^{58}$Department of Physics \& Astronomy, University of California Los Angeles, Los Angeles, CA 90095, USA\\
$^{59}$Department of Physics \& Astronomy, University of California Irvine, Irvine, CA 92697, USA\\
$^{60}$California Institute of Technology, Pasadena, CA 91125, USA\\
$^{61}$Patashnick Voorheesville Observatory, Voorheesville, NY 12186, USA\\
$^{62}$Institute of Planetary Research, German Aerospace Center, Rutherfordstrasse 2, 12489 Berlin, Germany\\
$^{63}$Hazelwood Observatory, Australia\\
$^{64}$Perth Exoplanet Survey Telescope, Perth, Western Australia\\
$^{65}$Earth and Planets Laboratory, Carnegie Institution for Science, 5241 Broad Branch Road NW, Washington, DC 20815, USA\\
$^{66}$NASA Hubble Fellow\\
$^{67}$Observatories of the Carnegie Institution for Science, 813 Santa Barbara Street, Pasadena, CA 91101, USA\\
$^{68}$Exoplanetary Science at UNSW, School of Physics, UNSW Sydney, NSW 2052, Australia\\
$^{69}$Departamento de Astronom\'ia, Universidad de Chile, Casilla 36-D, Santiago, Chile\\
$^{70}$School of Astronomy and Space Science, Key Laboratory of Modern Astronomy and Astrophysics in Ministry of Education, Nanjing University, Nanjing 210046, Jiangsu, China\\
%%%%%%%%%%%%%%%%%%%%%%%%%%%%%%%%%%%%%%%%%%%%%%%%%%

% Don't change these lines
\bsp	% typesetting comment
\label{lastpage}
\end{document}